\documentclass[paper,toc,nohyper]{JHEP3}
\usepackage{epsfig,cite}
\usepackage{amsbsy} 
\usepackage{epsfig}

\def\beq{\begin{equation}}   
\def\eeq{\end{equation}}
\def\bea{\begin{eqnarray}}  
\def\eea{\end{eqnarray}}

\def\O{y}

\def\CA{C_A}

\def\NF{N_F}

\def\d{\hbox{d}}

\def\ln{\hbox{ln}}

\title{\boldmath 
NNLO corrections to event shapes in $e^+e^-$ annihilation
}
\author{
A.~Gehrmann--De Ridder\\
Institute for Theoretical Physics, ETH, CH-8093 Z\"urich,
Switzerland\\ 
E-mail: \email{gehra@phys.ethz.ch}}
\author{
T.~Gehrmann\\
Institut f\"ur Theoretische Physik, Universit\"at Z\"urich,
Winterthurerstrasse 190,\\ CH-8057 Z\"urich, Switzerland\\
E-mail: \email{thomas.gehrmann@physik.unizh.ch}}
\author{E.W.N.~Glover\\
Institute of Particle Physics Phenomenology, 
        Department of Physics,\\
        University of Durham, Durham, DH1 3LE, UK\\
	E-mail: \email{e.w.n.glover@durham.ac.uk}}
\author{
G.~Heinrich\\
School of Physics, The University of Edinburgh, Edinburgh EH9 3JZ,
UK\\
E-mail: \email{gheinric@ph.ed.ac.uk}}

\abstract{
We compute the next-to-next-to-leading order (NNLO) QCD corrections to 
the six most important 
event shape variables related to three-particle final states 
in electron-positron annihilation. The corrections are sizeable for all
variables,  however their magnitude is substantially different for different
observables. We observe that the NNLO corrections yield a considerably 
better agreement between theory and experimental data both in shape and 
normalisation of the event shape distributions. The 
renormalisation scale dependence of the theoretical prediction
is substantially reduced compared to 
the previously existing NLO results. 
Our results will allow a precise determination of the strong coupling 
constant from event shape data collected at LEP. 
}
\preprint{{ZU-TH 27/07}, {IPPP/07/90}, Edinburgh 2007-47}

\begin{document}

\section{Introduction}
\label{sec:intro}

For more than a decade experiments at LEP (CERN) and SLC (SLAC) 
gathered  a wealth of high precision high energy hadronic data
from electron-positron annihilation at a range of centre-of-mass energies~\cite{aleph,delphi,l3,opal,sld}. 
This data provides one of the    
 cleanest
ways of probing our quantitative understanding of QCD. 
This is particularly so because the strong interactions occur only in 
the final state and are not entangled with the parton density functions associated 
with beams of hadrons.
As the understanding of the strong interaction, and the capability of 
making more precise theoretical predictions, develops, 
more and more stringent comparisons of theory and experiment are possible,
leading to improved measurements
of fundamental quantities such as the strong coupling constant~\cite{expreview}.

In addition
to measuring multi-jet production rates, more specific information  about the
topology of the events can be extracted. To this end, many variables  have been
introduced which characterise the hadronic structure of an event. For example,
we can ask how planar or how collimated an event is. In general, a variable is
described as $n$ jet-like if it vanishes for a final state configuration of
$(n-1)$ hadrons\footnote{It should be 
noted that sometimes in the literature, 
especially in works on resummation, event shapes requiring 
three particles are called two-jet event shapes, while those requiring 
four particles are called three-jet event shapes.}. 
With the precision data from LEP and SLC, experimental
distributions for such event shape variables have been extensively  studied and
have been compared with theoretical calculations based on next-to-leading order
(NLO)  parton-level event generator  programs~\cite{ERT,kunszt,event}, 
  improved by
resumming kinematically-dominant leading and next-to-leading logarithms
(NLO+NLL)~\cite{ctwt,ctw}  and by the inclusion of  
non-perturbative models of power-suppressed hadronisation
effects~\cite{power}. 

Comparing the different sources of error in the extraction of $\alpha_s$
from hadronic data,
one finds that the purely experimental error is negligible compared to
the theoretical uncertainty. There are two sources of theoretical
uncertainty: the theoretical description of the parton-to-hadron
transition (hadronisation uncertainty) and the uncertainty stemming from the 
truncation of the perturbative series at a certain order, as estimated by scale
variations (perturbative or scale uncertainty).  Although the precise
size of the hadronisation uncertainty is debatable and perhaps often
underestimated, it is conventional to consider the scale
uncertainty as the dominant source of theoretical error on the precise
determination of  $\alpha_s$ from three-jet observables. 

For the bulk of the paper we are concerned with the next-to-next-to-leading order (NNLO)
perturbative corrections to three jet-like shape variables. To be precise, we present the NNLO
coefficients for the differential distributions of thrust, the wide and total jet broadening,
heavy hemisphere mass, $C$ parameter  and the jet transition variable $Y_{3}$ for the Durham jet
algorithm. These results are obtained using a numerical implementation of the two-loop $\gamma^*
\to 3$~parton~\cite{3jme}, the one-loop $\gamma^* \to 4$~partons~\cite{V4p} and the tree-level
$\gamma^*\to 5$~parton matrix elements~\cite{tree5p}. Each of the contributions becomes singular
when one or more partons are soft and/or collinear. In previous work, we have developed an antenna subtraction
method~\cite{ourant} for isolating singularities and ensuring that the final result is infrared
finite~\cite{our3j}. The resulting numerical program, {\tt EERAD3}, yields the full kinematical
information on the partonic final state and can be applied to  generic infrared safe three-jet
observables.  

In section~\ref{sec:shapes}, we provide definitions of the relevant three-jet shape variables
while section~\ref{sec:PT} reviews the structure of the perturbative predictions.  
Section~\ref{sec:nnlo} gives a brief description of the NNLO calculation and its implementation in
the multi-purpose parton level Monte Carlo program {\tt EERAD3}. Results 
for the event
shape distributions are reported 
 in sections~\ref{sec:nnloresults} and \ref{sec:results},
together with an estimate of the remaining perturbative uncertainty due to
variations of the renormalisation scale. The parton level predictions are also
compared with hadron-level experimental data.  
Finally, our results are summarised in section~\ref{sec:conclusions}.

\section{Event shape variables}
\label{sec:shapes}

In order to characterise hadronic final states in electron-positron
annihilation, a variety of event shape variables have been proposed in 
the literature, for a review see e.g.~\cite{dissertori}. These variables can be categorised 
into two classes, according to the minimal number of final-state particles 
required for them to be non-vanishing: the most common variables require 
three particles (and are thus closely related to three-jet final states),
while several other variables were constructed such that they require 
at least four particles (related to four-jet final states). 

Among the event shapes requiring three-particle final states, 
six variables 
were studied in great detail: the thrust $T$~\cite{farhi}, the
normalised heavy jet mass $M_H^2/s$~\cite{mh}, 
the wide and total jet
broadenings $B_W$ and $B_T$~\cite{bwbt},  
the $C$-parameter~\cite{c} and the transition from three-jet to 
two-jet final states in the Durham jet algorithm $Y_3$~\cite{durham}.

\begin{itemize}
\item[(a)] Thrust, $T$~\cite{farhi}  \\
The thrust variable for a hadronic final state in $e^+e^-$ annihilation is
defined as~\cite{farhi}
\begin{equation}
T=\max_{\vec{n}}
\left(\frac{\sum_i |\vec{p_i}\cdot \vec{n}|}{\sum_i |\vec{p_i}|}\right)\,,
\label{eq:t}
\end{equation}
where $\vec{p_i}$ 
denotes the three-momentum of particle $i$, with the sum running
over all particles. The unit vector $\vec{n}$ is varied to find  the
thrust direction $\vec{n}_T$ which maximises the expression in parentheses. 

The maximum value of thrust, $T \to 1$, is obtained in the limit where there
are only two particles in the event. For a three-particle event the minimum
value of thrust is $T = 2/3$.

\item[(b)] Heavy hemisphere mass, $M_H^2/s$~\cite{mh} \\
In the original definition~\cite{mh} one divides the event into
two hemispheres.
In each hemisphere, $H_i$, one also computes the hemisphere invariant mass as:
\begin{equation}
M_i^2/s = \frac{1}{E_{{\rm vis}}^2} 
\left(\sum_{k\in H_i} p_k \right)^2\;,
\end{equation}
where $E_{{\rm vis}}$ is the total energy visible in the event. 
In the original definition, the hemisphere is chosen such that
$M_1^2+M_2^2$ is minimised.
We follow the more customary definition whereby the hemispheres
are separated by the plane orthogonal to the thrust axis.

The larger of the two hemisphere invariant masses yields the  
heavy jet mass:
\begin{equation}
\rho \equiv M_H^2/s = {\rm max}( M_1^2/s , M_2^2/s )\;. 
\end{equation}
In the two-particle limit $\rho \to 0$, while for a 
three-particle event   
$\rho \leq {1}/{3}$.

The associated light hemisphere mass,
\begin{equation}
M_L^2/s = {\rm min}( M_1^2/s , M_2^2/s )  
\end{equation}
is an example of a four-jet observable and vanishes in the 
three-particle limit.

At lowest order, the heavy jet mass and the $(1-T)$ distribution are
identical.   However, this degeneracy is lifted at next-to-leading order.

\item[(c)] Jet Broadening, $B_W$ and $B_T$~\cite{bwbt}\\
Taking a plane perpendicular to $\vec{n}_T$ through
the coordinate origin, one defines two event hemispheres $H_{1,2}$. In 
each of them, one determines the hemisphere broadening:
\begin{equation}
B_{i} = \frac{{\displaystyle \sum_{k\in H_i} | \vec{p_k} \times \vec{n}_T |}}
{{\displaystyle 2 \sum_k |\vec{p_k}|}}\,. 
\end{equation}
The wide and total jet broadening are then defined as
\begin{eqnarray}
B_W &=& {\rm max}(B_1,B_2)\;, \label{eq:bw}\\
B_T &=& B_1+B_2 \;. \label{eq:bt}
\end{eqnarray}
In the two-particle limit $B_W \to 0$ and $B_T \to 0$.
The maximum broadening for a three-particle event is 
$B_T = B_W = {1}/(2\sqrt{3})$.

The narrow jet broadening,
\begin{equation}
B_N = {\rm min}(B_1,B_2)\;, \label{eq:bn}
\end{equation}
is another four-jet observable and vanishes when only three particles are
in the event.

\item[(d)] The $C$ parameter, \cite{c}\\

The linearised momentum tensor
\begin{equation}
\Theta^{\alpha\beta} = \frac{1}{\sum_k |\vec{p_k}|} \, 
\sum_{k}\frac{p_k^\alpha p_k^\beta}{ |\vec{p_k}|}\,, \qquad 
(\alpha,\beta = 1,2,3)\;,
\end{equation} 
has three eigenvalues $\lambda_i$, which are used to construct the 
$C$-parameter:
\begin{equation}
C= 3\, \left( \lambda_1\lambda_2 + \lambda_2\lambda_3 + \lambda_3\lambda_1
\right) \;.
\label{eq:c}
\end{equation}
This definition is equivalent to 
\begin{equation}
C = 3\, \left( \Theta^{11}\Theta^{22} +  \Theta^{22}\Theta^{33}
+ \Theta^{33}\Theta^{11} - \Theta^{12}\Theta^{12} 
- \Theta^{23}\Theta^{23}  - \Theta^{31}\Theta^{31}  
\right)\;.
\end{equation}

The related four-jet observable is the $D$-parameter,
\begin{equation}
D= 27\, \lambda_1\lambda_2 \lambda_3  \;.
\label{eq:d}
\end{equation}

\item[(e)] The jet transition variable, $Y_3$~\cite{durham}\\

The jet transition variable $Y_3$ is defined as the value of the jet
resolution parameter $y_{{\rm cut}}$ for which an event changes from a
three-jet to a two-jet configuration with some
jet defining scheme. 

Here, we focus on the Durham jet algorithm which clusters particles into jets 
by computing the measurement variable
\begin{equation}
y_{ij,D} = \frac{2 \, {\rm min} (E_i^2,E_j^2) (1-\cos \theta_{ij})}
{E_{{\rm vis}}^2}
\end{equation}
for each pair ($i,j$) of particles. The pair with the lowest 
$y_{ij,D}$ is replaced by a pseudoparticle whose four-momentum is 
given by the 
sum of the four-momenta of particles $i$ and $j$ ('E' recombination 
scheme). This procedure is repeated as long as pairs with invariant 
mass below the predefined resolution parameter
$y_{ij,D}<y_{{\rm cut}}$  are found. Once the clustering is terminated, the 
remaining (pseudo-)particles are the jets. 

\end{itemize}

\section{Event shapes in perturbation theory}
\label{sec:PT}

The perturbative expansion for the distribution of a 
generic observable $\O$ up to NNLO at centre-of-mass energy $\sqrt{s}$ 
for renormalisation scale $\mu^2 = s$ and 
$\alpha_s\equiv \alpha_s(\sqrt s)$  is given by
\begin{eqnarray}
\frac{1}{\sigma_{{\rm had}}}\, \frac{\d\sigma}{\d \O} &=& 
\left(\frac{\alpha_s}{2\pi}\right) \frac{\d \bar A}{\d \O} +
\left(\frac{\alpha_s}{2\pi}\right)^2 \frac{\d \bar B}{\d \O}
+ \left(\frac{\alpha_s}{2\pi}\right)^3 
\frac{\d \bar C}{\d \O} + {\cal O}(\alpha_s^4)\;.
\label{eq:NNLO}
\end{eqnarray}
Here the event shape distribution  
is normalised to the total hadronic cross section $\sigma_{\rm{had}}$.
With the assumption of massless quarks, then   
at NNLO 
we have,
  \begin{equation}
  \sigma_{\rm{had}}=\sigma_0\,
\left(1+\frac{3}{2}C_F\,\left(\frac{\alpha_s}{2\pi}\right)
+K_2\,\left(\frac{\alpha_s}{2\pi}\right)^2+{\cal O}(\alpha_s^3)\,
\right) \;,
\end{equation}
where the Born cross section for $e^+e^- \to q \bar q$ is
\begin{equation}
\sigma_0 = \frac{4 \pi \alpha}{3 s} N \, e_q^2\;.
\end{equation}
The constant $K_2$ is given by,
\begin{equation}
  K_2=\frac{1}{4}\left[- \frac{3}{2}C_F^2
+C_FC_A\,\left(\frac{123}{2}-44\zeta_3\right)+C_FT_RN_F\,(-22+16\zeta_3)
 \right] \;,
\end{equation}
where the QCD colour factors are
\begin{equation}
\CA = N,\qquad C_F = \frac{N^2-1}{2N},
\qquad T_R = \frac{1}{2}\; 
\end{equation}
for $N=3$ colours and $N_F$ light quark flavours.

In practice, we compute the perturbative coefficients $A$, $B$ and $C$, which are 
all normalised to 
$\sigma_0$:
\begin{eqnarray}
\frac{1}{\sigma_0}\, \frac{\d\sigma}{d \O} &=& 
\left(\frac{\alpha_s}{2\pi}\right) \frac{\d  A}{\d \O} +
\left(\frac{\alpha_s}{2\pi}\right)^2 \frac{\d  B}{\d \O}
+ \left(\frac{\alpha_s}{2\pi}\right)^3 
\frac{\d  C}{\d \O} + {\cal O}(\alpha_s^4)\,.
\label{eq:NNLOsigma0}
\end{eqnarray}
However, $A$, $B$ and $C$ are straightforwardly related to $\bar{A}$, $\bar{B}$ 
and $\bar{C}$,
\begin{eqnarray}
\bar{A} &=& A\;,\nonumber \\
\bar{B} &=& B - \frac{3}{2}C_F\,A\;,\nonumber \\
\bar{C} &=& C -  \frac{3}{2}C_F\,B+ \left(\frac{9}{4}C_F^2\,-K_2\right)\,A 
\;.\label{eq:ceff}
\end{eqnarray} 
These coefficients are computed at a renormalisation scale fixed to 
the centre-of-mass energy, 
  and 
depend therefore only on the value of the observable $y$.
They explicitly include only QCD corrections with non-singlet 
quark couplings and are therefore independent of electroweak 
couplings.
At ${\cal O}(\alpha_s^2)$, these amount to the full 
corrections, while the ${\cal O}(\alpha_s^3)$ corrections also 
receive a pure-singlet contribution. This pure-singlet contribution 
arises from the interference of diagrams where the external gauge boson 
couples to different quark lines. In four-jet observables at 
 ${\cal O}(\alpha_s^3)$, these singlet contributions were found to be 
extremely small~\cite{dixonsigner}.
Also, the pure-singlet contribution from three-gluon final states 
to three-jet observables was found to be  negligible~\cite{nigeljochum}. 
This small correction to NNLO is denoted by $\delta_C$:
\begin{equation}
\frac{1}{\sigma_0}\,
\frac{\d \sigma}{\d y} \Bigg|_{{\rm NNLO,\, pure\; singlet}}
=\left(\frac{\alpha_s}{2\pi}\right)^3 \, 
\frac{\d \delta_C}{\d y}(s,M_Z,\alpha,\sin^2\Theta_W, c_q) 
\end{equation}
where $c_q$ denotes the set of all electroweak vector and axial-vector 
quark couplings.

First-order electroweak corrections 
to event shape observables could be of a magnitude 
comparable to the NNLO QCD corrections. Like the pure-singlet NNLO 
contributions, these do also not factorise onto $\sigma_0$. 
The first-order electroweak corrections affect the distribution itself 
and the normalisation $\sigma_{{\rm had}}$. Collectively, they 
give a contribution
of the form,
\begin{equation}
\frac{1}{\sigma_{{\rm had}}}\,
\frac{\d \sigma}{\d y} \Bigg|_{{\rm electroweak}, {\cal O}(\alpha\alpha_s)}
= \left(\frac{\alpha}{2\pi}\right) \left(\frac{\alpha_s}{2\pi}\right) 
\, \frac{\d \delta_{EW}}{\d y}(s,M_Z,\alpha,\sin^2\Theta_W, c_q).
\end{equation}
These corrections are not complete at present~\cite{moretti}, 
and clearly deserve further 
study.

In summary, the expression for event shape distributions accurate to 
NNLO in QCD and NLO in the electroweak theory reads:
\begin{eqnarray}
\frac{1}{\sigma_{{\rm had}}}\, \frac{\d\sigma}{\d \O} &=& 
\left(\frac{\alpha_s}{2\pi}\right) \frac{\d \bar A }{\d \O} +
\left(\frac{\alpha_s}{2\pi}\right)^2 \frac{\d \bar B }{\d \O}
+ \left(\frac{\alpha_s}{2\pi}\right)^3 
\frac{\d \bar C }{\d \O} \nonumber \\ &&
+ \left(\frac{\alpha_s}{2\pi}\right)^3  
\frac{\d \delta_C}{\d \O}(s,M_Z,\alpha,\sin^2\Theta_W, c_q) 
\nonumber \\ &&+ \left(\frac{\alpha}{2\pi}\right)
\left(\frac{\alpha_s}{2\pi}\right) 
\frac{\d \delta_{EW}}{\d \O}(s,M_Z,M_H,\alpha,\sin^2\Theta_W, c_q)\;.
\end{eqnarray}
In the following, we will focus on the QCD non-singlet expression 
(\ref{eq:NNLO}), since $\delta_C$ can be safely neglected, and 
the computation of $\delta_{EW}$ needs further work.

The QCD coupling constant evolves according to the renormalisation group 
equation, which is to NNLO:
\begin{equation}
\label{eq:running}
\mu^2 \frac{\d \alpha_s(\mu)}{\d \mu^2} = -\alpha_s(\mu) 
\left[\beta_0 \left(\frac{\alpha_s(\mu)}{2\pi}\right) 
+ \beta_1 \left(\frac{\alpha_s(\mu)}{2\pi}\right)^2 
+ \beta_2 \left(\frac{\alpha_s(\mu)}{2\pi}\right)^3 
+ {\cal O}(\alpha_s^4) \right]\,
\end{equation}
with the $\overline{{\rm MS}}$-scheme coefficients
\begin{eqnarray}
\beta_0 &=& \frac{11 \CA - 4 T_R \NF}{6}\;,\nonumber  \\
\beta_1 &=& \frac{17 \CA^2 - 10 C_A T_R \NF- 6C_F T_R \NF}{6}\;, \nonumber \\
\beta_2 &=&\frac{1}{432}
\big( 2857 C_A^3 + 108 C_F^2 T_R N_F -1230 C_FC_A T_R N_F
-2830 C_A^2T_RN_F \nonumber \\ &&
+ 264 C_FT_R^2 N_F^2 + 316 C_AT_R^2N_F^2\big)\;.
\end{eqnarray}

Equation~(\ref{eq:running}) 
is solved by introducing $\Lambda$ as integration constant
with $L= \log(\mu^2/\Lambda^2)$, yielding the running coupling constant:
\begin{equation}
\alpha_s(\mu) = \frac{2\pi}{\beta_0 L}\left( 1- 
\frac{\beta_1}{\beta_0^2}\, \frac{\log L}{L} + \frac{1}{\beta_0^2 L^2}\,
\left( \frac{\beta_1^2}{\beta_0^2}\left( \log^2 L - \log L - 1
\right) + \frac{\beta_2}{\beta_0}  \right) \right)\;.
\end{equation}

In terms of the running coupling $\alpha_s(\mu)$, the 
NNLO (non-singlet) expression for event shape distributions becomes
\begin{eqnarray}
\frac{1}{\sigma_{{\rm had}}}\, \frac{\d\sigma}{\d \O} (s,\mu^2,\O) &=& 
\left(\frac{\alpha_s(\mu)}{2\pi}\right) \frac{\d \bar A}{\d \O} +
\left(\frac{\alpha_s(\mu)}{2\pi}\right)^2 \left( 
\frac{\d \bar B}{\d \O} + \frac{\d \bar A}{\d \O} \beta_0 
\log\frac{\mu^2}{s} \right)
\nonumber \\ &&
+ \left(\frac{\alpha_s(\mu)}{2\pi}\right)^3 
\bigg(\frac{\d \bar C}{\d \O} + 2 \frac{\d \bar B}{\d \O}
 \beta_0\log\frac{\mu^2}{s}
+ \frac{\d \bar A}{\d \O} \left( \beta_0^2\,\log^2\frac{\mu^2}{s}
+ \beta_1\, \log\frac{\mu^2}{s}   \right)\bigg)  
\nonumber \\ &&
 + {\cal O}(\alpha_s^4)\;.
\label{eq:NNLOmu} 
\end{eqnarray}

\section{Calculation of NNLO corrections}
\label{sec:nnlo}

Three-jet production at tree-level is induced by the decay of a virtual
photon (or other neutral gauge boson) into a quark-antiquark-gluon final
state. At higher orders, this process receives corrections from extra
real or virtual particles. The individual partonic channels
that contribute through to NNLO
are shown in Table~\ref{table:partons}. All of the tree-level and loop
amplitudes associated with these channels are known 
in the literature~\cite{3jme,muw2,V4p,tree5p}.

\begin{table}[th]
\begin{center}
\begin{tabular}{lll}
\hline\\
LO & $\gamma^*\to q\,\bar qg$ & tree level \\[2mm]
NLO & $\gamma^*\to q\,\bar qg$ & one loop \\
 & $\gamma^*\to q\,\bar q\, gg$ & tree level \\
 & $\gamma^*\to q\,\bar q\, q\bar q$ & tree level \\[2mm]
NNLO & $\gamma^*\to q\,\bar qg$ & two loop \\
 & $\gamma^*\to q\,\bar q\, gg$ & one loop \\
& $\gamma^*\to q\,\bar q\, q\,\bar q$ & one loop \\
& $\gamma^*\to q\,\bar q\, q\,\bar q\, g$ & tree level \\
& $\gamma^*\to q\,\bar q\, g\,g\,g$ & tree level\\[2mm]
\hline
\end{tabular}
\caption{Non-singlet partonic contributions to three-jet event shape
observables in perturbative QCD.\label{table:partons}}
\end{center}
\end{table}

For a given partonic final state, the event shape observable $\O$ 
is computed according to 
the same definition as in the experiment, which is applied to partons instead 
of hadrons. At leading order, all three final state partons must be 
well separated from each other, such that $\O$ differs from the trivial 
two-parton limit. At NLO, up to four partons 
can be present in the final state, 
two of which 
can be clustered together,   
whereas at NNLO, the final state can consist of up to five partons, 
and as many as three partons can be clustered together. 
The more partons in the final state, 
the better one expects the matching between theory and 
experiment to be~\cite{whynnlo}.

The two-loop $\gamma^* \to q\bar q g$ matrix elements were derived 
in~\cite{3jme} by reducing all relevant Feynman integrals to a small 
set of master integrals using integration-by-parts~\cite{ibp} and 
Lorentz invariance~\cite{gr} identities, solved with the Laporta 
algorithm~\cite{laporta}. The master integrals~\cite{3jmaster} were 
computed from their differential equations~\cite{gr} and expressed 
analytically
in terms of one- and two-dimensional harmonic polylogarithms~\cite{hpl}. 

The one-loop four-parton matrix elements relevant here~\cite{V4p} were 
originally derived in the context of NLO corrections to four-jet 
production and related event shapes~\cite{fourjetprog,cullen}. One of 
these four-jet parton-level event 
generator programs~\cite{cullen} is the starting point 
for our calculation, since it already contains all relevant 
four-parton and five-parton matrix elements.

The four-parton and five-parton contributions to three-jet-like final 
states at NNLO contain infrared real radiation singularities, which 
have to be extracted and combined with the 
infrared singularities~\cite{catani} 
present in the virtual three-parton and four-parton contributions to 
yield a finite result. In our case, this is accomplished by 
introducing subtraction functions, which account for the 
infrared real radiation singularities, and are sufficiently simple to 
be integrated analytically. Schematically, this subtraction reads:
\begin{eqnarray*}
\lefteqn{{\rm d}\sigma_{NNLO}=\int_{{\rm d}\Phi_{5}}\left({\rm d}\sigma^{R}_{NNLO}
-{\rm d}\sigma^{S}_{NNLO}\right) }\nonumber \\ 
&&+\int_{{\rm d}\Phi_{4}}\left({\rm d}\sigma^{V,1}_{NNLO}
-{\rm d}\sigma^{VS,1}_{NNLO}\right)\nonumber \\&&
+ \int_{{\rm d}\Phi_{5}}
{\rm d}\sigma^{S}_{NNLO}
+\int_{{\rm d}\Phi_{4}}{\rm d}\sigma^{VS,1}_{NNLO} 
+ \int_{{\rm d}\Phi_{3}}{\rm d}\sigma^{V,2}_{NNLO}\;,
\end{eqnarray*}
where ${\rm d} \sigma^{S}_{NNLO}$ denotes the real radiation subtraction term 
coinciding with the five-parton tree level cross section 
 ${\rm d} \sigma^{R}_{NNLO}$ in all singular limits~\cite{doubleun}. 
Likewise, ${\rm d} \sigma^{VS,1}_{NNLO}$
is the one-loop virtual subtraction term 
coinciding with the one-loop four-parton cross section 
 ${\rm d} \sigma^{V,1}_{NNLO}$ in all singular limits~\cite{onelstr}. 
Finally, the two-loop correction 
to the three-parton cross section is denoted by ${\rm d}\sigma^{V,2}_{NNLO}$.
With these, each line in the above equation is individually 
infrared finite, and 
can be integrated numerically.

Systematic methods to derive and integrate subtraction terms 
were available in the literature only to NLO~\cite{nlosub,ant}.
Physical results for the special case
  of NNLO Higgs production
  have been achieved in~\cite{cshiggs}.
In the context of 
this project, we fully developed an NNLO subtraction formalism~\cite{our2j,ourant,our3j}, 
based on the antenna subtraction method originally proposed at 
NLO~\cite{cullen,ant}. 
The basic idea of the antenna subtraction approach is to construct 
the subtraction terms  from antenna functions. 
Each antenna function encapsulates 
all singular limits due to the 
 emission of one or two unresolved partons between two colour-connected hard
partons.
This construction exploits the universal factorisation of 
phase space and squared matrix elements in all unresolved limits.
The individual antenna functions are obtained by normalising 
three-parton and four-parton tree-level matrix elements and 
three-parton one-loop matrix elements 
to the corresponding two-parton 
tree-level matrix elements. Three different types of 
antenna functions are required,
corresponding to the different pairs of hard partons 
forming the antenna: quark-antiquark, quark-gluon and gluon-gluon antenna 
functions. All these can be derived systematically from matrix 
elements~\cite{our2j} for physical processes. 

The factorisation of the final state phase space into antenna phase 
space and hard phase space requires a mapping of the antenna momenta 
onto reduced hard momenta. We use the mapping derived in~\cite{dak1} for 
the three-parton and four-parton antenna functions. To extract the infrared 
poles of the subtraction terms, the antenna functions must be integrated 
analytically over the appropriate antenna phase spaces, which is done by 
reduction~\cite{babis} to known 
phase space master integrals~\cite{ggh}. 

We tested the proper implementation of 
the subtraction by generating trajectories of phase space points approaching 
a given single or double unresolved limit. 
Along these trajectories, we observe that the 
antenna subtraction terms converge towards the physical matrix 
elements, and that the cancellations among individual 
contributions to the subtraction terms take place as expected. 
Moreover, we checked the correctness of the 
subtraction by introducing a 
lower cut (slicing parameter) on the phase space variables, and observing 
that our results are independent of this cut (provided it is 
chosen small enough). This behaviour indicates that the 
subtraction terms ensure that the contribution of potentially singular 
regions of the final state phase space does not contribute to the numerical 
integrals, but is accounted for analytically. A detailed description of the 
calculation can be found in~\cite{our3j}. 

The resulting numerical program, {\tt EERAD3}, yields the full kinematical 
information on a given multi-parton final state. It 
can thus be used to compute any 
infrared-safe observable related to three-particle final states at 
${\cal O}(\alpha_s^3)$ in $e^+e^-$-annihilation. 

\section{NNLO distributions}
\label{sec:nnloresults}

In this section, we discuss the size and shape of the  LO, NLO and NNLO coefficients
of the perturbative expansion of the various event shape observables defined in 
Eq.~(\ref{eq:NNLOsigma0}).   
For convenience, we weight the distribution by the 
observable.   

The precise size and shape of the NNLO corrections depend on the observable in
question.  However, all contributions are dominated by the behaviour in the two-jet
region where the observable generally tends to zero. Of course,  typical hadronic events
contain many hadrons and it is extremely unlikely that the value of any
event shape is precisely zero for any experimental event. However, in the fixed
order partonic calculation, where there are at most five partons present in the final 
state, one or more of the partons may be soft and/or collinear,  and the observable
may approach zero. In such circumstances, soft gluon singularities cause the fixed
order prediction to become wildly unstable and grow logarithmically. 
In the infrared
limit $\O \to 0$, the perturbative coefficients have the following form,
\begin{eqnarray}
\O\frac{\d A}{\d \O} &\sim& A_{1}L  + A_{0}\nonumber \\
\O\frac{\d B}{\d \O} &\sim& B_{3}L^3 +B_{2}L^2 + B_{1}L   +B_{0}\nonumber \\
\O\frac{\d C}{\d \O} &\sim& C_{5}L^5 +C_{4}L^4 + C_{3}L^3 +C_{2}L^2 +C_{1}L + C_{0}
\end{eqnarray}
where $L = \ln(1/\O)$ and $C_{n}$ are (as yet) undetermined coefficients. 
Whenever $L$ is sufficiently large, resummation effects will be important.  
In our numerical studies, we therefore
impose a cut on the size of $\O$ which is typically in the range 0.001 -- 0.01,
since for such small values of $\O$ we do not trust the fixed order prediction.

Even away from the infrared region, the shape of the fixed order prediction is 
heavily influenced by cancellations between the real and virtual contributions.  The
LO contribution $A$ is very large and positive at small $\O$ and decreases
monotonically as $\O$  increases. The NLO contribution $B$ is negative at small
$\O$, but exhibits a turn-over, typically at $\O \sim 0.05$.    Similarly,  the NNLO
contribution $C$  also exhibits a turn-over, but at a slightly larger value of $\O$.
The precise positions of the maxima of the distributions depend on the observable
under consideration.

A second generic feature occurs when  the paucity of final state particles imposes
a maximum value for the observable.   Examples include $(1-T)$ and $C$ which are 
required to be less than $0.33$ and $0.75$ respectively for three-parton final
states.   As the number of partons increases with the perturbative order, this limit
is relaxed and larger values of the observable are accessed.

Finally, typical values of the strong coupling constant lie around $\alpha_s \sim
0.12$, so that $\frac{\alpha_s}{2\pi} \sim 1/50$. 
 It is well known that the NLO
corrections are large,  $B_{\O} \sim (15-30) A_{\O}$, leading to a  30-60\% NLO
effect in the region where the perturbative calculation is expected to be reliable.
However, we observe that in all cases,  the NNLO  coefficients are also significant,
$C_{\O} \sim (200-800) A_{\O}$, leading to a further 7-28\% NNLO correction.

\subsection{Thrust}

\begin{figure}[t]
\begin{center}
\epsfig{file=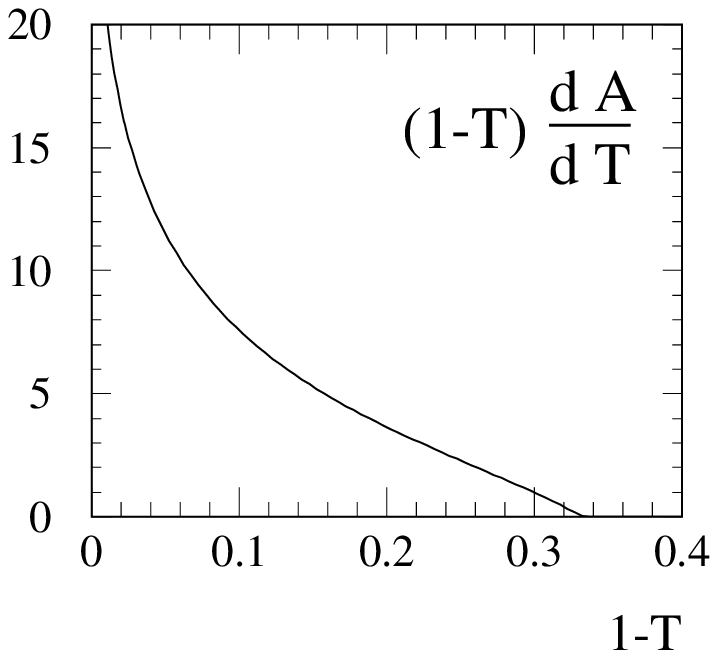,angle=-90,width=4.5cm}
\epsfig{file=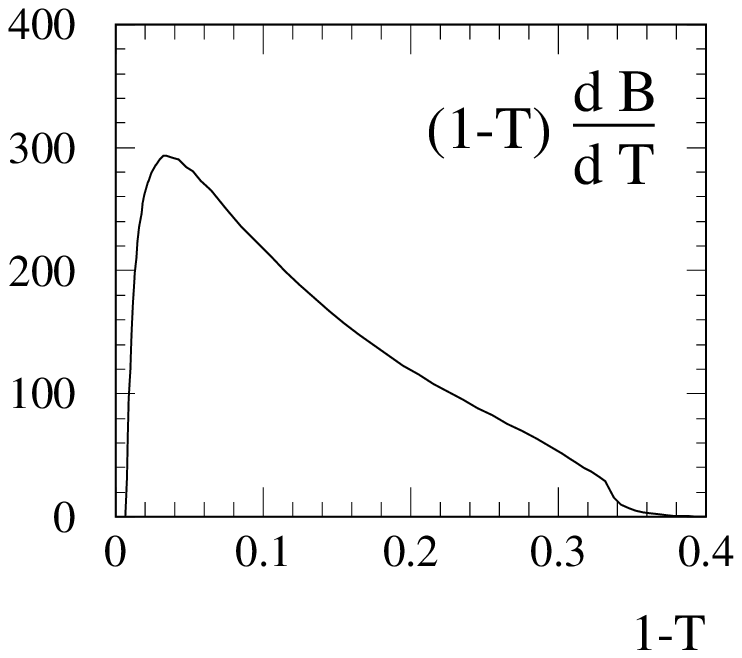,angle=-90,width=4.5cm}
\epsfig{file=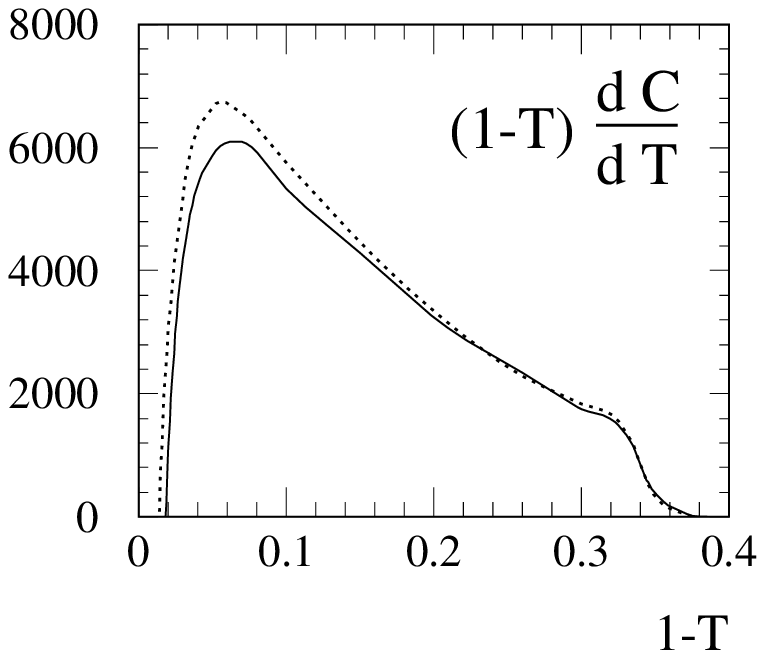,angle=-90,width=4.5cm}
\end{center}
\caption{Coefficients of the leading order, next-to-leading order and
next-to-next-to-leading order 
contributions to the thrust distribution as defined in 
Eq.~(\protect{\ref{eq:NNLOsigma0}}) and weighted by $(1-T)$.
The dotted line in the $C$ coefficient indicates the distribution prior to 
correction of the soft large-angle radiation terms (see erratum at the end of the paper).
}\label{fig:thrust-abc}
\end{figure}

Thrust is defined in section~\ref{sec:shapes}(a).  First results for the NNLO corrections to the thrust
distribution were presented in Ref.~\cite{ourthrust}. The perturbative coefficients for the thrust
distribution weighted by $(1-T)$ are shown in Fig.~\ref{fig:thrust-abc}.    As discussed earlier,
the shape of the contribution is dominated by the infrared region at $(1-T) \to 0$.    At small
$(1-T)$, the LO contribution $A$ is very large and positive, while the NLO and NNLO coefficients
$B$ and $C$  are rising 
and exhibit a turn-over at moderate values of $(1-T)$.   We observe that the
peak moves from about 0.04 (NLO) to 0.06 (NNLO).   We also see that the NLO and NNLO
distributions progressively extend to  larger and larger values of $(1-T)$ as the phase space
restrictions on  large values of $(1-T)$ are relaxed.  In the intermediate region, $0.04 < (1-T)
< 0.33$,  we observe that the perturbative coefficients are roughly in the ratio,   $A:B:C
~\sim 1:30:800$. Setting $\alpha_s\sim 0.12$ 
and using Eq.~(\ref{eq:ceff}), this indicates corrections which are of 
relative magnitude 
LO~:~NLO~:~NNLO $\sim 1:0.53:0.27$, such that the NNLO corrections increase the 
NLO prediction by another 18\%.

\subsection{Heavy jet mass}

\begin{figure}[t]
\begin{center}
\epsfig{file=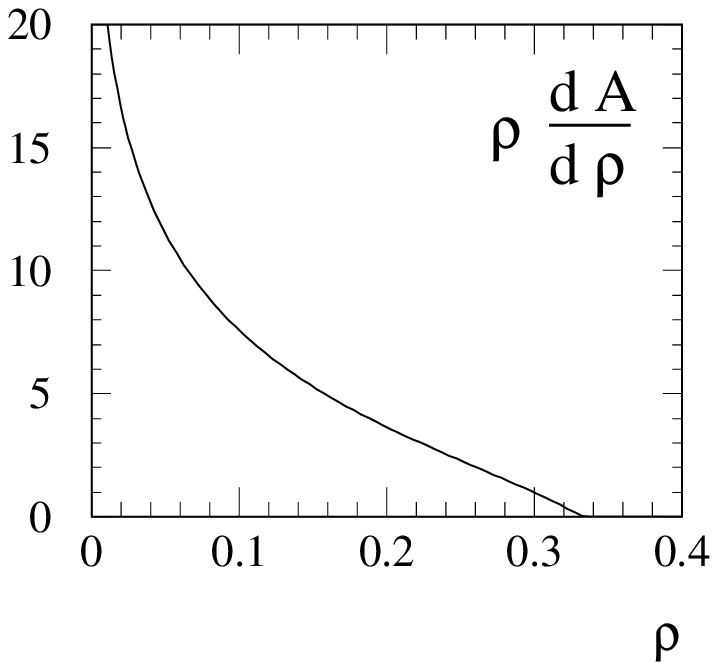,angle=-90,width=4.5cm}
\epsfig{file=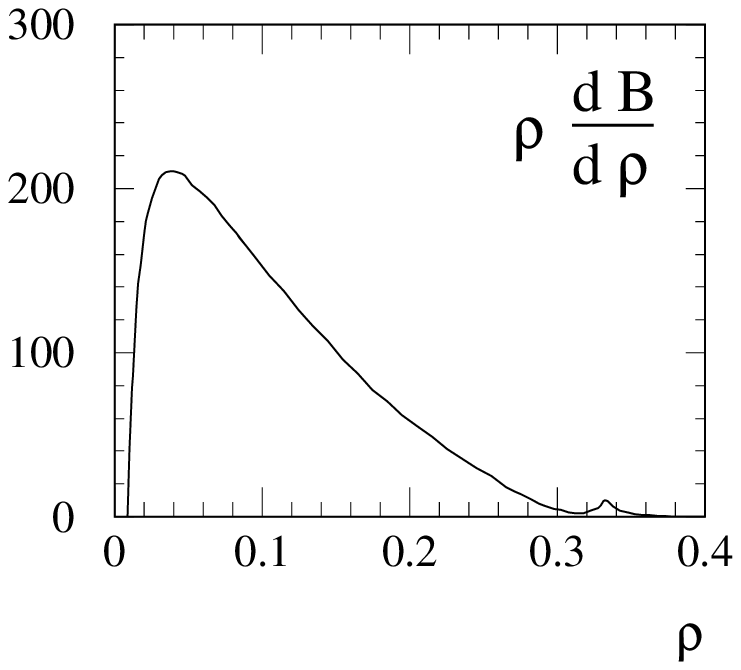,angle=-90,width=4.5cm}
\epsfig{file=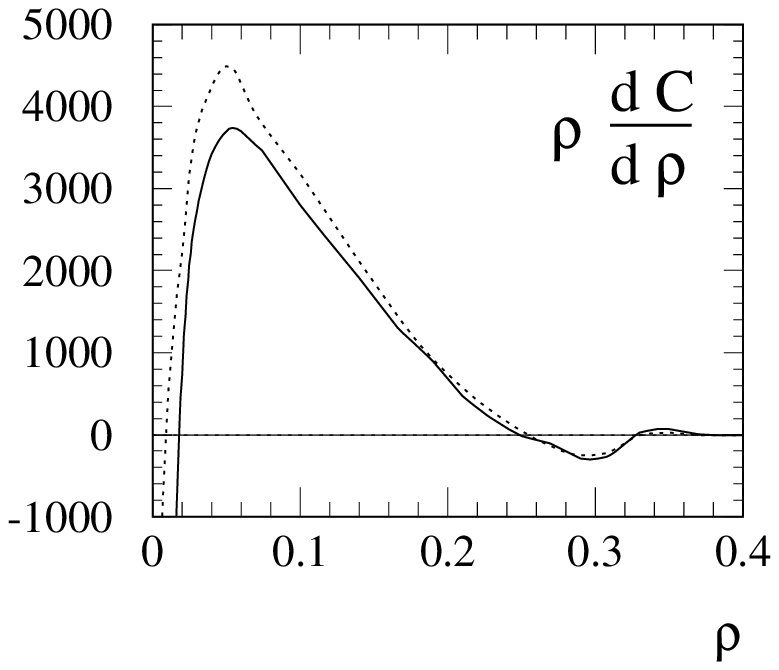,angle=-90,width=4.5cm}
\end{center}
\caption{Coefficients of the leading order, next-to-leading order and
next-to-next-to-leading order 
contributions to the heavy jet mass distribution as defined in 
Eq.~(\protect{\ref{eq:NNLOsigma0}}) and weighted by $\rho$.
The dotted line in the $C$ coefficient indicates the distribution prior to 
correction of the soft large-angle radiation terms.}\label{fig:mh-abc}
\end{figure}

The definition of the heavy jet mass given in section~\ref{sec:shapes}(b)   is the  larger invariant mass
of the two hemispheres formed by separating  the event by a plane normal to the thrust axis.  
The perturbative coefficients for the heavy jet mass distribution weighted by $\rho$  are shown
in Fig.~\ref{fig:mh-abc}. At lowest order, the heavy jet mass and the $(1-T)$ distribution are
identical, so that $A$ does not extend past $\rho = 0.33$.   At higher orders, the distribution
extends to larger values, with a small negative NNLO  contribution around $0.33$.    In the
intermediate region, $0.02 < \rho < 0.33$, the perturbative coefficients are roughly  $A:B:C
~\sim 1:20:400$ indicating corrections of approximately  
LO~:~NLO~:~NNLO $\sim 1:0.34:0.13$, translating into 
a 10\% enhancement of NNLO over NLO.
Comparing Fig.~\ref{fig:thrust-abc}(b) with \ref{fig:mh-abc}(b) and Fig.~\ref{fig:thrust-abc}(c)
with  \ref{fig:mh-abc}(c) we see clearly the rather different 
behaviour of the higher order
corrections to these observables,  
particularly in the region beyond the LO kinematic bound where
partonic configurations with two or more partons in each hemisphere 
contribute differently to each observable.

\subsection{Jet broadenings}
\label{sec:broad}

\begin{figure}[t]
\begin{center}
\epsfig{file=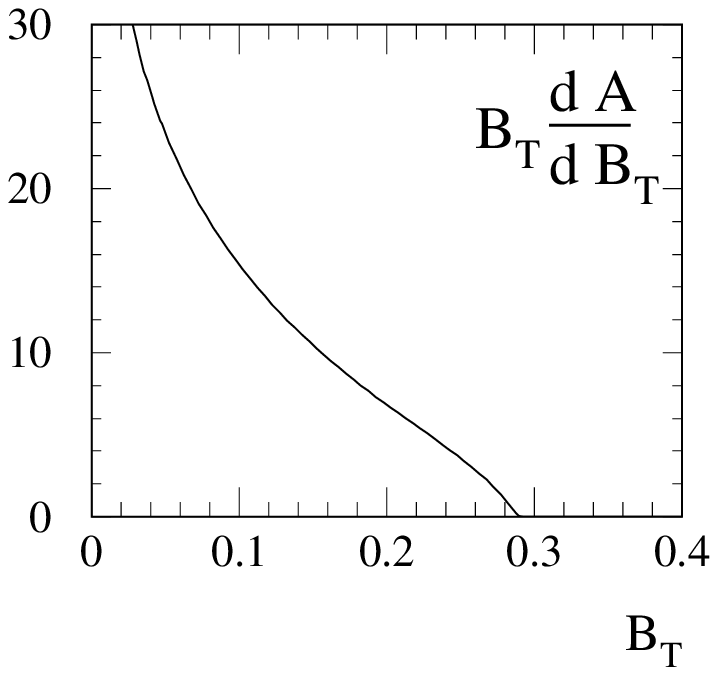,angle=-90,width=4.5cm}
\epsfig{file=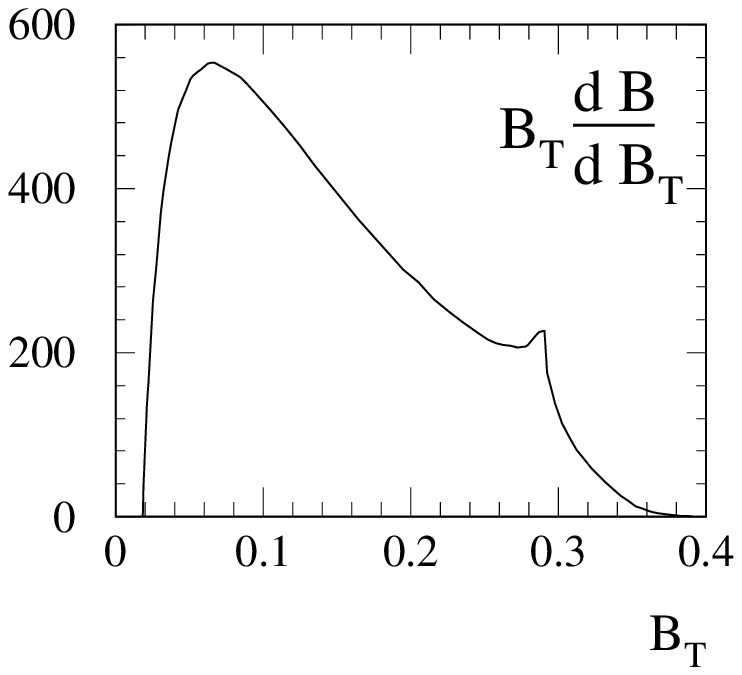,angle=-90,width=4.5cm}
\epsfig{file=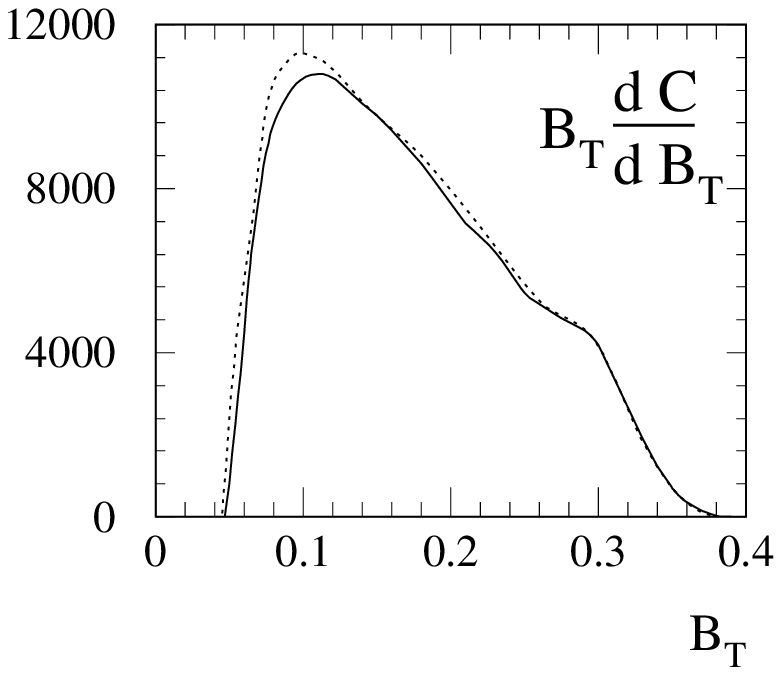,angle=-90,width=4.5cm}
\end{center}
\caption{Coefficients of the leading order, next-to-leading order and
next-to-next-to-leading order 
contributions to the total jet broadening distribution as defined in 
Eq.~(\protect{\ref{eq:NNLOsigma0}}) and weighted by $B_T$. The dotted line in the $C$ coefficient indicates the distribution prior to 
correction of the soft large-angle radiation terms.}\label{fig:bt-abc}
\end{figure}

\begin{figure}[t]
\begin{center}
\epsfig{file=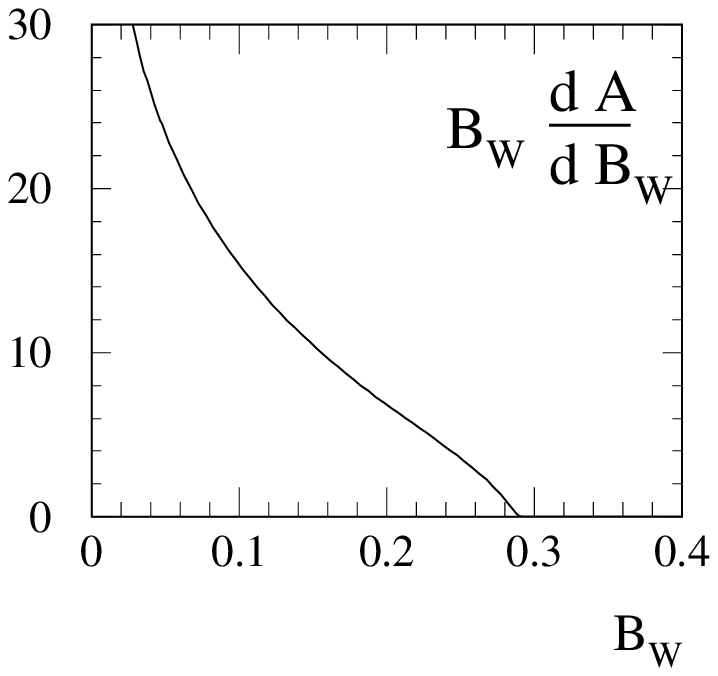,angle=-90,width=4.5cm}
\epsfig{file=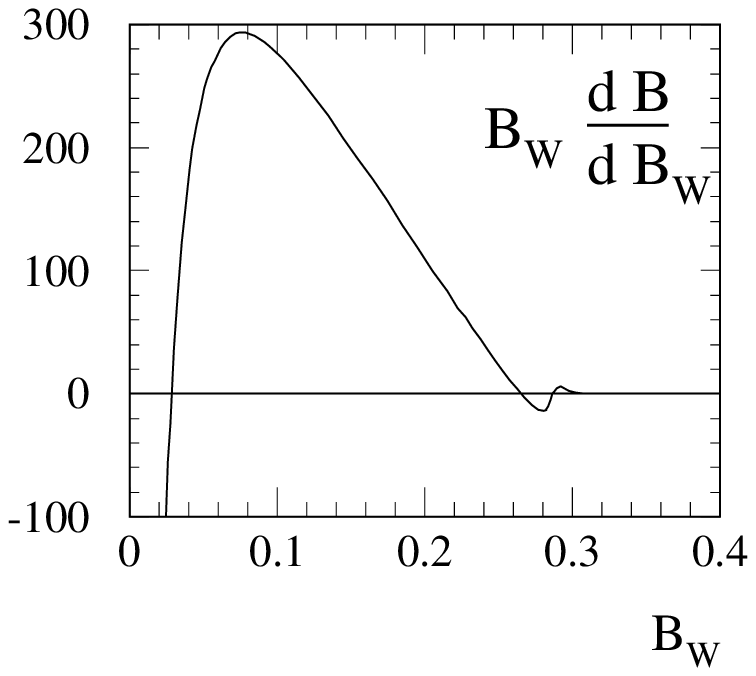,angle=-90,width=4.5cm}
\epsfig{file=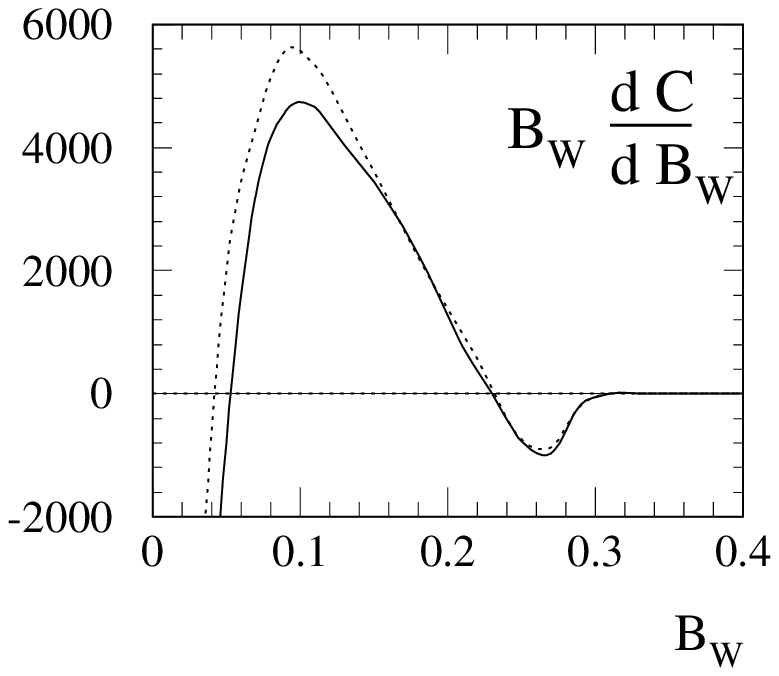,angle=-90,width=4.5cm}
\end{center}
\caption{Coefficients of the leading order, next-to-leading order and
next-to-next-to-leading order 
contributions to the wide jet broadening distribution as defined in 
Eq.~(\protect{\ref{eq:NNLOsigma0}}) and weighted by $B_W$. The dotted line in the $C$ coefficient indicates the distribution prior to 
correction of the soft large-angle radiation terms.}\label{fig:bw-abc}
\end{figure}

The jet broadenings are defined in section~\ref{sec:shapes}(c) by dividing the event into two
hemispheres using a plane normal to the thrust axis..    At lowest order $B_W$ and $B_T$ are
identical, but their distributions receive different higher order corrections from partonic
configurations with two or more partons in each hemisphere.

The perturbative coefficients for the $B_T$ ($B_W$) distributions weighted by
$B_T$ ($B_W$) are shown in Fig.~\ref{fig:bt-abc} (Fig.~\ref{fig:bw-abc})
respectively. The structures evident around $B_T,~B_W\sim (1/2\sqrt{3}) 
\sim 0.29$ are generated
by four and five parton events and are therefore different for the 
two observables.
For more moderate $B_T$ values between 0.04 and 0.29, the
perturbative coefficients are in the ratio $A:B:C ~\sim 1:35:800$.  Including
the factors of $\alpha_s$, this  leads 
to corrections LO~:~NLO~:~NNLO $\sim
1:0.63:0.27$  for $\alpha_s\sim 0.12$, which amounts to NNLO corrections of 
17\% of the NLO result.       
   We observe that the corrections for
$B_W$ are 
considerably 
smaller than those for $B_T$, $A:B:C ~\sim 1:20:400$ or equivalently
$LO:NLO:NNLO \sim 1:0.34:0.13$, which yields
a 10\% NNLO effect over NLO.   When $B_T < 0.04$ ($B_W< 0.04$), 
infrared logarithms must be
resummed to produce a meaningful result.

\subsection{$C$-parameter}

\begin{figure}[t]
\begin{center}
\epsfig{file=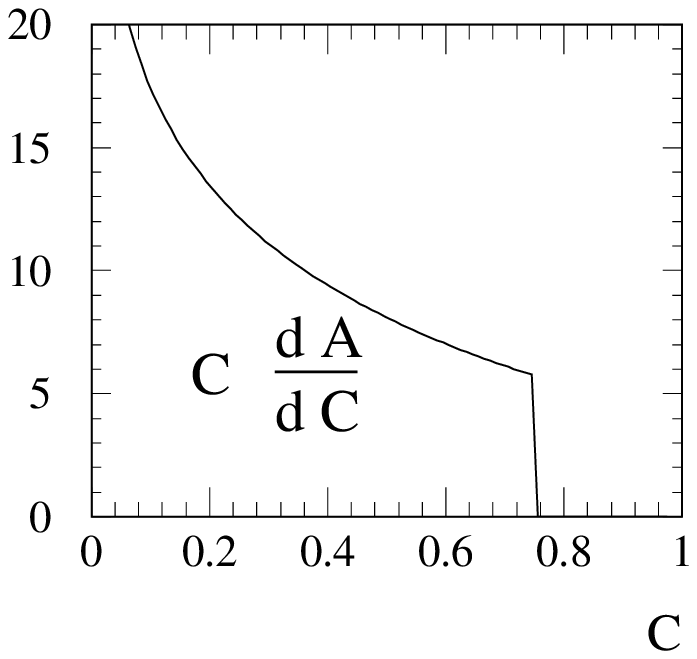,angle=-90,width=4.5cm}
\epsfig{file=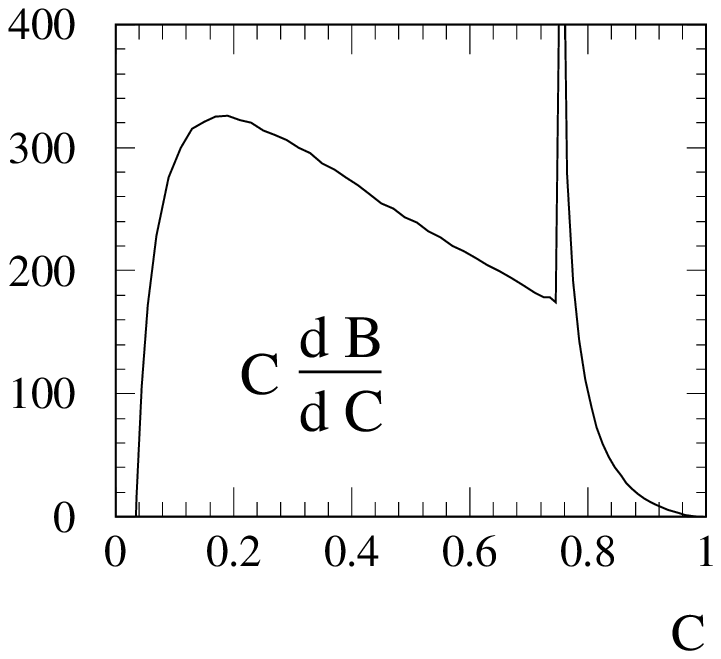,angle=-90,width=4.5cm}
\epsfig{file=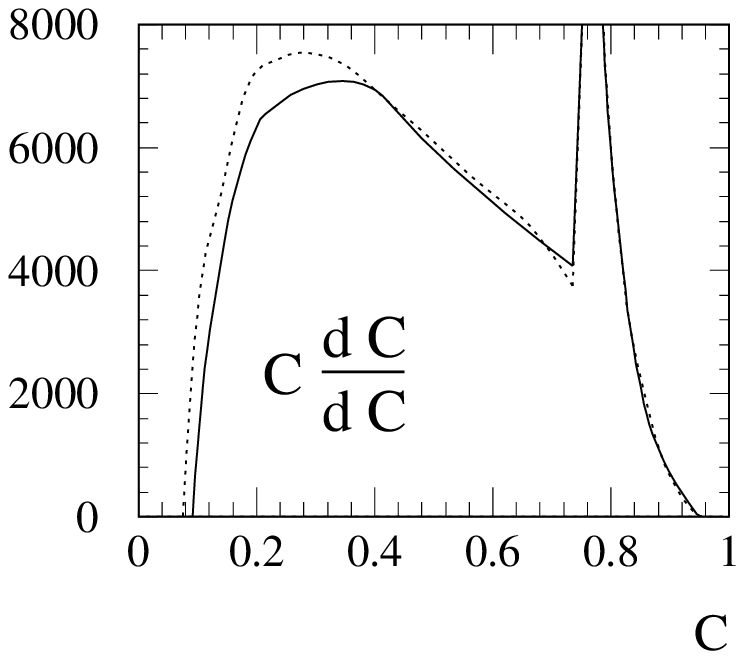,angle=-90,width=4.5cm}
\end{center}
\caption{Coefficients of the leading order, next-to-leading order and
next-to-next-to-leading order 
contributions to the $C$ parameter distribution as defined in 
Eq.~(\protect{\ref{eq:NNLOsigma0}}) and weighted by $C$. The dotted line in the $C$ coefficient indicates the distribution prior to 
correction of the soft large-angle radiation terms.}\label{fig:c-abc}
\end{figure}

The $C$ parameter is defined in section~\ref{sec:shapes}(d) and the perturbative
distributions at LO, NLO and NNLO weighted by $C$  are shown in
Fig.~\ref{fig:c-abc}. The LO kinematic limit at $C=0.75$ is clearly visible. 
At NLO (and NNLO), four (and five) parton events can generate larger values of
$C$,  leading to a sharp peak around $C~\sim 0.75$. The approximate size of the
corrections for $0.1 < C < 0.75$ is $A:B:C ~\sim 1:30:700$, 
or, including the
factors of $(\alpha_s/2\pi)$ with $\alpha_s\sim 0.12$, in the ratio
LO~:~NLO~:~NNLO $\sim 1:0.53:0.23$, resulting in 
a 15\% enhancement of NNLO over NLO.  At smaller values of $C < 0.1$, 
large infrared logarithms render the fixed order prediction unreliable and must be
resummed.  Similarly, large logarithms are produced around the LO kinematic limit,
$C\sim 0.75$ which must also be resummed.

\subsection{$Y_3$}

\begin{figure}[t]
\begin{center}
\epsfig{file=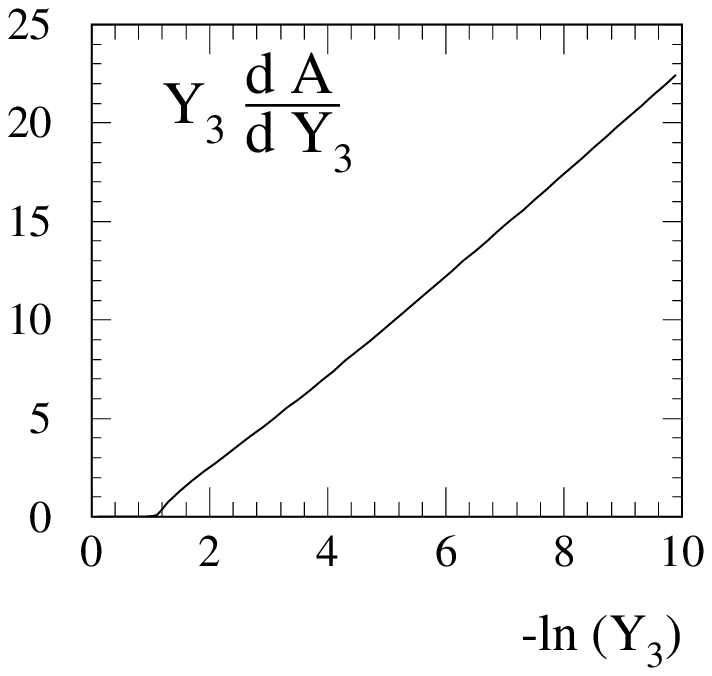,angle=-90,width=4.5cm}
\epsfig{file=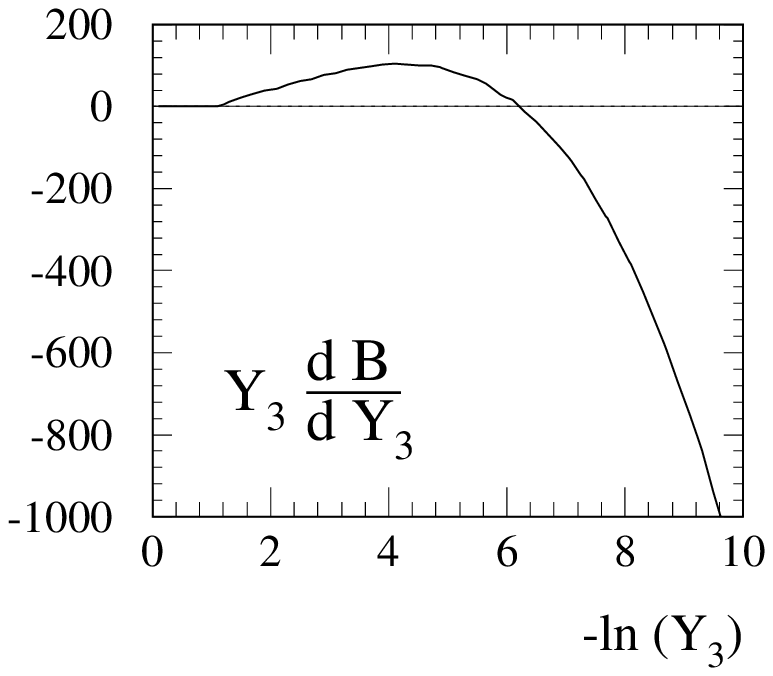,angle=-90,width=4.5cm}
\epsfig{file=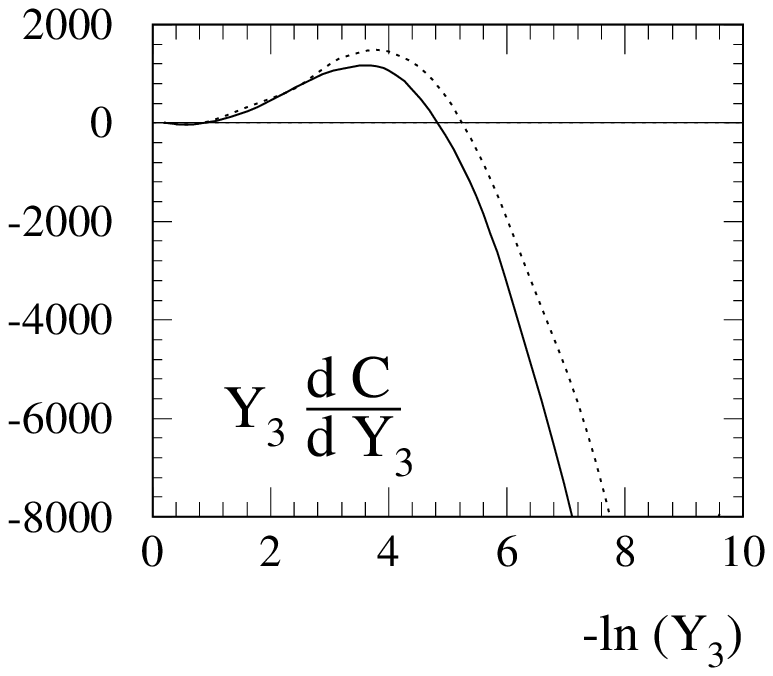,angle=-90,width=4.5cm}
\end{center}
\caption{Coefficients of the leading order, next-to-leading order and
next-to-next-to-leading order 
contributions to the  distribution of the jet transition variable $Y_3$ as defined in 
Eq.~(\protect{\ref{eq:NNLOsigma0}}) and weighted by $Y_3$. The dotted line in the $C$ coefficient indicates the distribution prior to 
correction of the soft large-angle radiation terms.}\label{fig:y23-abc}
\end{figure}

The jet transition variable $Y_3$ is defined in section~\ref{sec:shapes}(e). It
describes the value of the jet resolution parameter $y_{{\rm cut}}$ for
which an event changes from a three-jet to a two-jet configuration within the
Durham jet algorithm. The perturbative distributions at LO, NLO and
NNLO weighted by $Y_3$  are shown in Fig.~\ref{fig:c-abc}.  As with all 
of the event shapes,
$Y_3\d A/\d Y_3$ is linear when plotted on a logarithmic scale.
For moderate values of $Y_3$, $2 < -\ln(Y_3) < 6$, the corrections are positive.  In this region,
the approximate size of the
corrections is $A:B:C ~\sim 1:15:200$, or, including the
factors of $(\alpha_s/2\pi)$ with $\alpha_s\sim 0.12$, in the ratio
LO~:~NLO~:~NNLO $\sim 1:0.25:0.06$, which produces a 5\% NNLO effect over NLO. 
However, at
smaller values of $Y_3$ (larger values of -$\ln(Y_3)$) 
resummation of logarithmic contributions are clearly mandatory.

\section{Comparison with data}
\label{sec:results}

We have presented the NNLO corrections to six event-shape distributions.
As we have shown in the previous section, the magnitude of the NNLO correction is
different for the six variables. 

Each of the event shapes considered here has been studied in depth by all four
experiments at LEP at centre-of-mass  energies of 91.2, 133, 161, 172, 183, 189, 200
and 206 GeV~\cite{aleph,delphi,l3,opal}. Within the experimental uncertainties, 
these data sets are mutually consistent.   

In this paper, we select data from ALEPH~\cite{aleph} as a representative set of
hadronic final states in electron-positron annihilation to illustrate the
improvement in the theoretical prediction due to the inclusion of the NNLO
perturbative contribution. The only free parameter in our
predictions is the strong coupling constant; we use the current 
world average value $\alpha_s(M_Z) = 0.1189$~\cite{bethke}.  

The experimental event-shape distributions were computed using the reconstructed
momenta and energies of charged and neutral particles. The measurements have been
corrected for detector effects and the final distributions correspond to the
particle (or hadron) level (stable hadrons and leptons after
hadronisation). In addition, at LEP2 energies above the $Z$ peak the data were corrected
for initial-state radiation effects and backgrounds, mainly from $W$-pair production,
were subtracted. The experimental uncertainties were estimated by varying event and
particle selection cuts and are below 1\% at LEP1 and between 0.5\% and 1.5\% at
LEP2. For further details we refer the interested reader to Ref.~\cite{aleph}.

\subsection{Thrust}

\begin{figure}[t]
\begin{center}
\epsfig{file=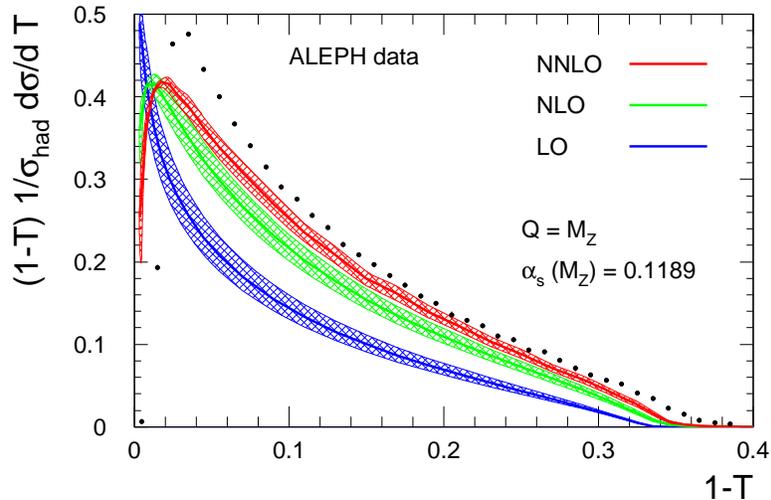,angle=-90,width=10cm}
\end{center}
\caption{Thrust distribution at $Q= M_Z$ at LO (blue), NLO (green) and NNLO
(red). The solid lines represent the prediction for renormalisation scale 
$\mu=Q$ and $\alpha_s(M_Z) = 0.1189$, while the
shaded region shows the variation due to varying
the renormalisation scale between $\mu=Q/2$ and $\mu = 2 Q$.
The data is taken from~\protect{\cite{aleph}}.\label{fig:thrust}}
\end{figure}
\begin{figure}[t]
\begin{center}
\epsfig{file=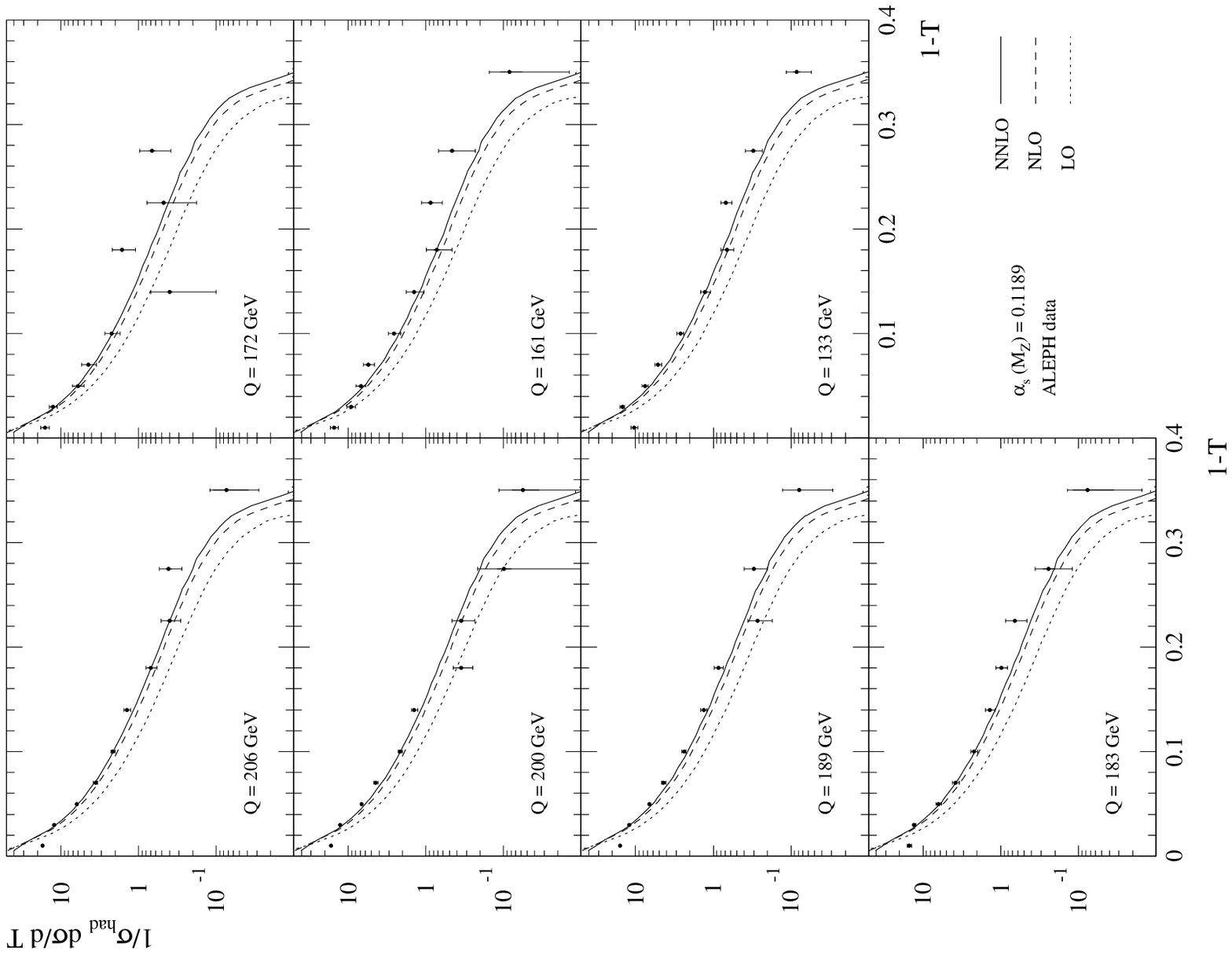,angle=-90,width=15cm}
\end{center}
\caption{The thrust distribution (with $\mu=Q$ and $\alpha_s(M_Z) = 0.1189$)
at LO (dotted), NLO (dashed) and NNLO
(solid) compared to experimental 
data from ALEPH~\protect{\cite{aleph}} for $Q = 133~{\rm GeV},\ldots,206$~GeV.}
\label{fig:aleph-thrust}
\end{figure}

Figure~\ref{fig:thrust} displays the perturbative  expression for the thrust distribution\footnote{First results for the NNLO corrections to the thrust
distribution were presented in Ref.~\cite{ourthrust}} at LO,
NLO and NNLO, evaluated  at $Q= M_Z$. The error band indicates the variation of the  prediction
under shifts of the renormalisation scale  in the range $\mu \in [Q/2;2\,Q]$ around the $e^+e^-$
centre-of-mass  energy $Q$.  The relative scale uncertainty   is reduced by about 30\% between
NLO and NNLO. 

The inclusion of the NNLO corrections enhances the thrust distribution by around (15-20)\% over the 
range $0.04 < (1-T) < 0.33$, where $-\ln(1-T)$ is not too large.  Outside this range, one does
not expect the  perturbative fixed-order prediction to yield reliable results.  For $(1-T)\to
0$,  the convergence of the perturbative series  is spoilt by powers of logarithms $\ln(1-T)$
appearing in higher perturbative orders,  thus necessitating an all-order resummation of these
logarithmic  terms~\cite{ctwt,ctw}, and a matching of fixed-order and resummed 
predictions~\cite{hasko}.

The perturbative parton-level prediction is compared with the hadron-level data from
the ALEPH collaboration~\cite{aleph} in Figure~\ref{fig:thrust} and 
Figure~\ref{fig:aleph-thrust}.  We observe that for all $Q$ values,
the shape and normalisation of the parton level NNLO prediction agrees better with
the data than at NLO. We also see that the NNLO corrections account for
approximately half of the difference between the parton-level NLO prediction and the hadron-level data.  

\subsection{Heavy jet mass}

\begin{figure}[t]
\begin{center}
\epsfig{file=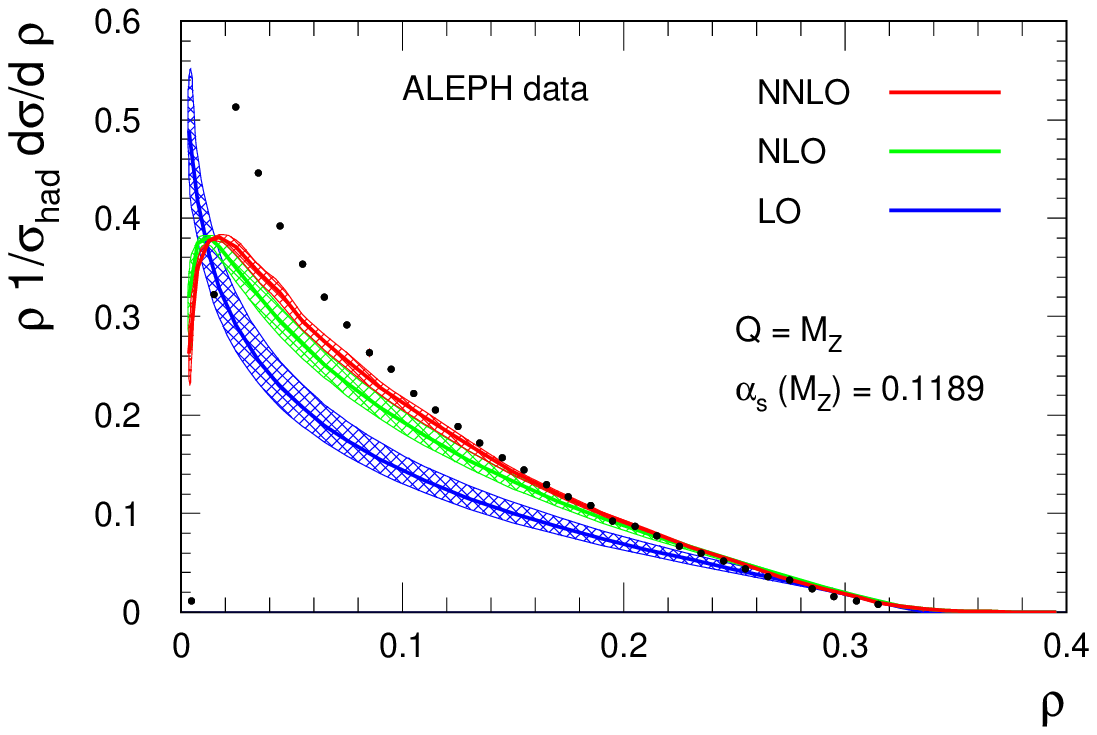,angle=-90,width=12cm}
\end{center}
\caption{Heavy jet mass distribution at $Q= M_Z$ at LO (blue), NLO (green) and NNLO
(red). The solid lines represent the prediction for renormalisation scale 
$\mu=Q$ and $\alpha_s(M_Z) = 0.1189$, while the
shaded region shows the variation due to varying
the renormalisation scale between $\mu=Q/2$ and $\mu = 2 Q$.
The data is taken from~\protect{\cite{aleph}}.\label{fig:mh}}
\end{figure}
\begin{figure}[t]
\begin{center}
\epsfig{file=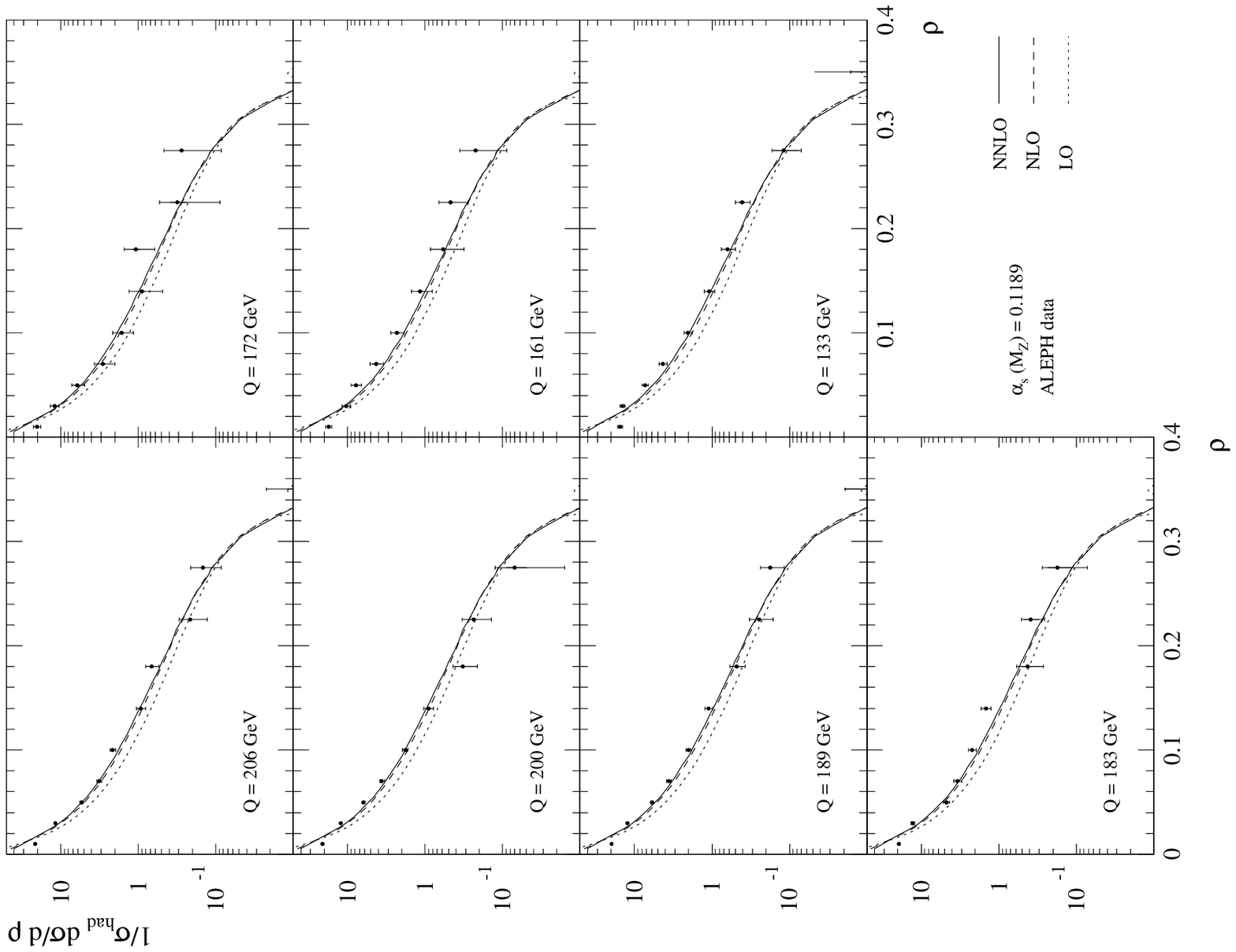,angle=-90,width=15cm}
\end{center}
\caption{Heavy jet mass distribution (with $\mu=Q$ and $\alpha_s(M_Z) = 
0.1189$)
at LO (dotted), NLO (dashed) and NNLO
(solid) compared to experimental 
data from ALEPH~\protect{\cite{aleph}} for $Q = 133~{\rm GeV},\ldots,206$~GeV.
\label{fig:aleph-mh}}
\end{figure}

The perturbative prediction for the heavy jet mass distribution is displayed in 
Figure~\ref{fig:mh}.   The solid lines represent the prediction at the  physical scale $Q= M_Z$,
while the shaded bands represent the effect of varying the renormalisation scale upwards and
downwards by a factor of 2. We observe that the relative scale uncertainty is reduced by about
50\% between NLO and NNLO.  It is noteworthy that the original motivation 
for introducing the heavy jet mass distribution~\cite{mh} was the
hope for  
improved perturbative stability over the thrust distribution. This improved 
stability was not evident from the existing  NLO results alone, but
becomes visible at NNLO. 

Compared to NLO, the inclusion of the NNLO corrections enhances 
the heavy jet mass distribution
by around 10\% over the  range $0.02 < \rho < 0.33$,  
where $\ln(\rho)$ is not too
large.  At smaller $\rho$ values, 
large $\ln(1/\rho)$ corrections must be resummed to all
orders~\cite{resum-mh} and matched onto the perturbative 
prediction. Nevertheless, in the
moderate to large $\rho$ region, the NNLO corrections render the fixed  order prediction significantly
closer to the experimental data~\cite{aleph}.

Figure~\ref{fig:aleph-mh} shows the prediction for a range of
 $Q$ values together  with the
hadron-level data from the ALEPH collaboration~\cite{aleph}. 
 For this observable, the NNLO
corrections are relatively small, however, for all $Q$ values, 
the shape and normalisation of
the parton-level NNLO prediction agrees slightly  
better with the hadron-level data than at NLO.

\subsection{Jet broadenings}
\begin{figure}[t]
\begin{center}
\epsfig{file=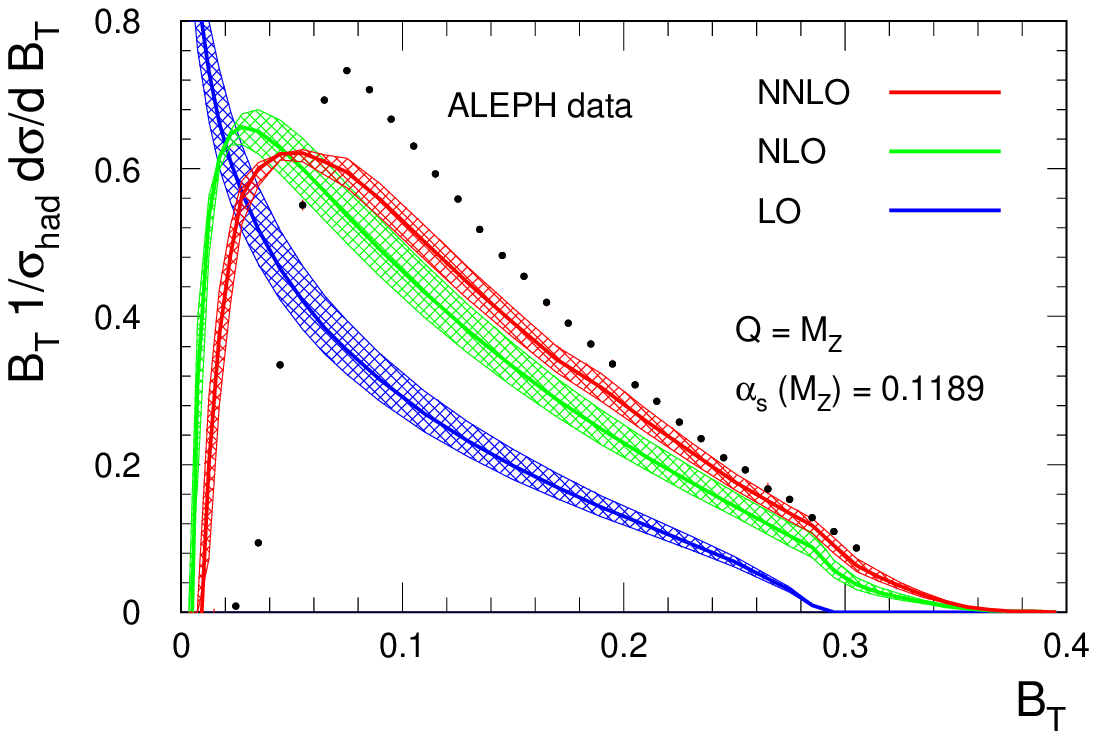,angle=-90,width=12cm}
\end{center}
\caption{Total jet broadening distribution at $Q= M_Z$ at LO (blue), NLO (green) and NNLO
(red). The solid lines represent the prediction for renormalisation scale 
$\mu=Q$ and $\alpha_s(M_Z) = 0.1189$, while the
shaded region shows the variation due to varying
the renormalisation scale between $\mu=Q/2$ and $\mu = 2 Q$.
The data is taken from \protect{\cite{aleph}}.\label{fig:bt}}
\begin{center}
\epsfig{file=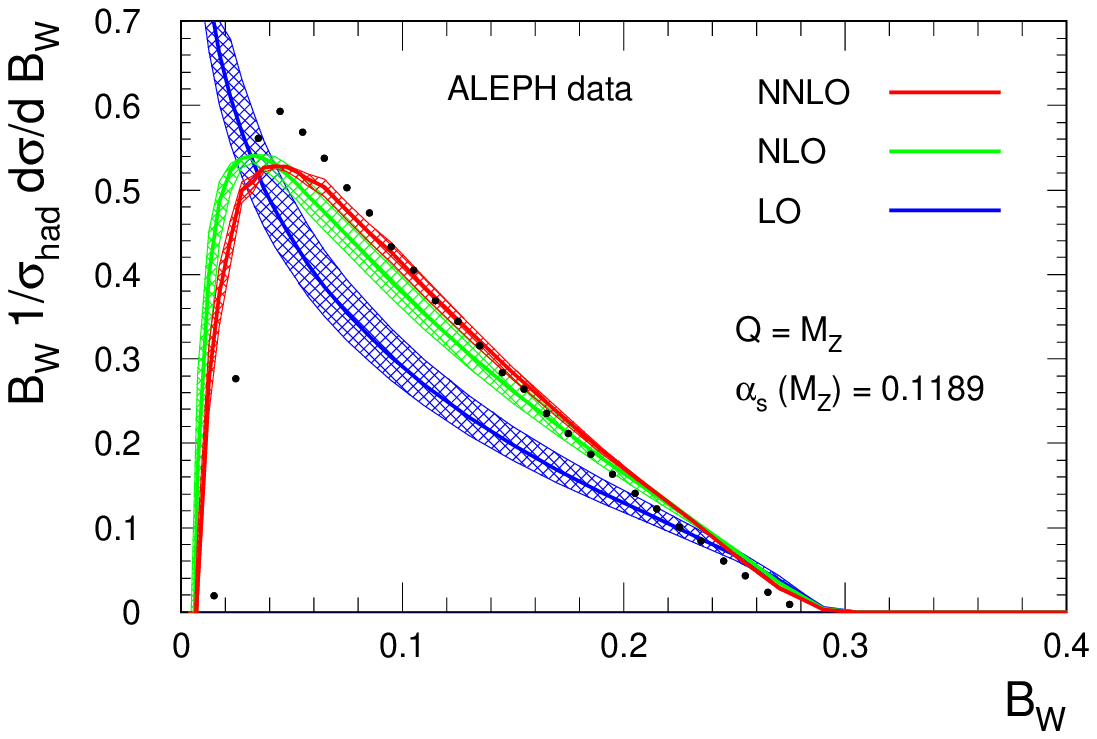,angle=-90,width=12cm}
\end{center}
\caption{Wide jet broadening distribution at $Q= M_Z$ at LO (blue), NLO (green) and NNLO
(red). The solid lines represent the prediction for renormalisation scale 
$\mu=Q$ and $\alpha_s(M_Z) = 0.1189$, while the
shaded region shows the variation due to varying
the renormalisation scale between $\mu=Q/2$ and $\mu = 2 Q$.
The data is taken from \protect{\cite{aleph}}.\label{fig:bw}}
\end{figure}
\begin{figure}[t]
\begin{center}
\epsfig{file=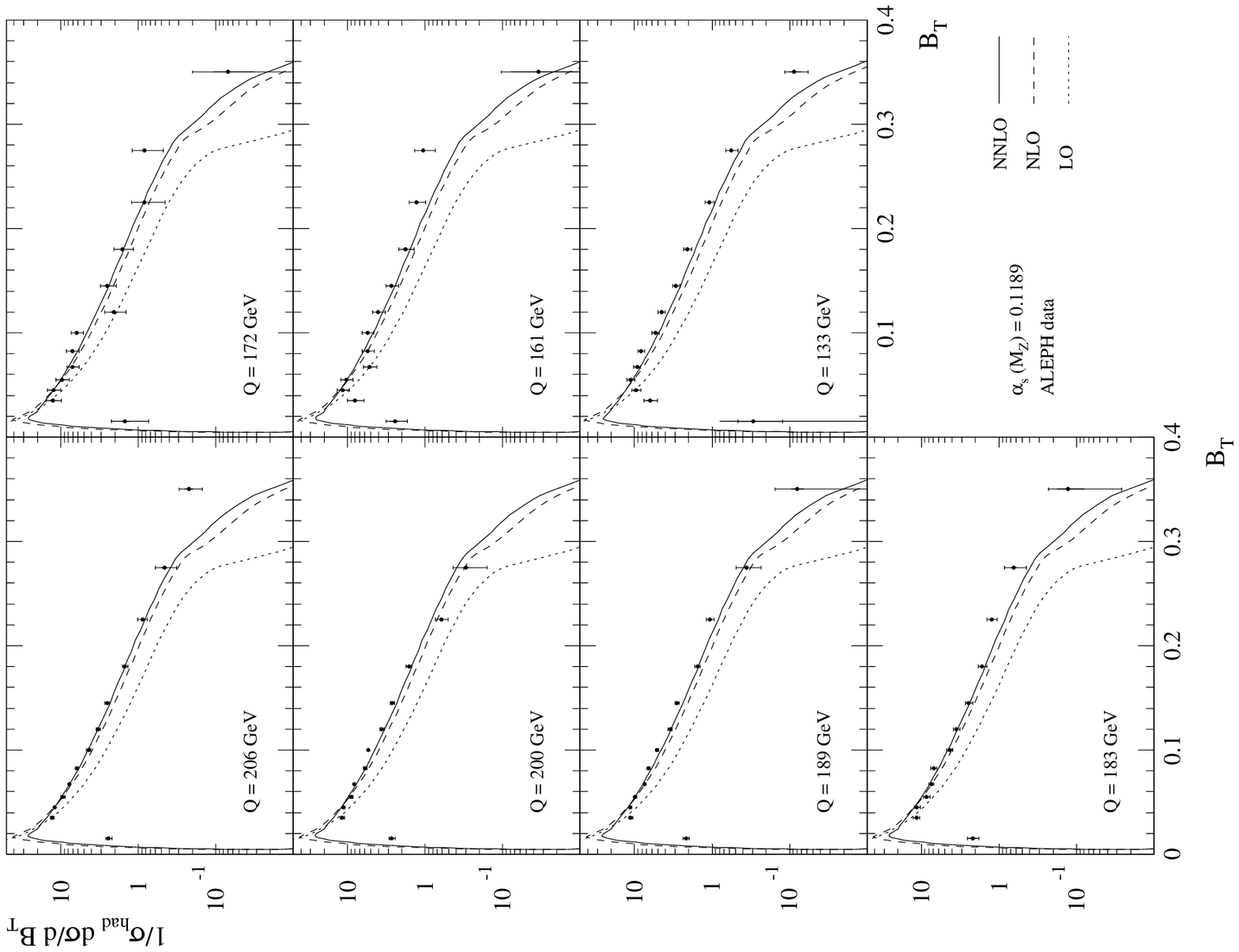,angle=-90,width=15cm}
\end{center}
\caption{Total jet broadening distribution (with $\mu=Q$ and $\alpha_s(M_Z) = 0.1189$)
at LO (dotted), NLO (dashed) and NNLO
(solid) compared to experimental 
data from ALEPH~\protect{\cite{aleph}} for $Q = 133~{\rm GeV},\ldots,206$~GeV.
\label{fig:aleph-bt}}
\end{figure}
\begin{figure}[t]
\begin{center}
\epsfig{file=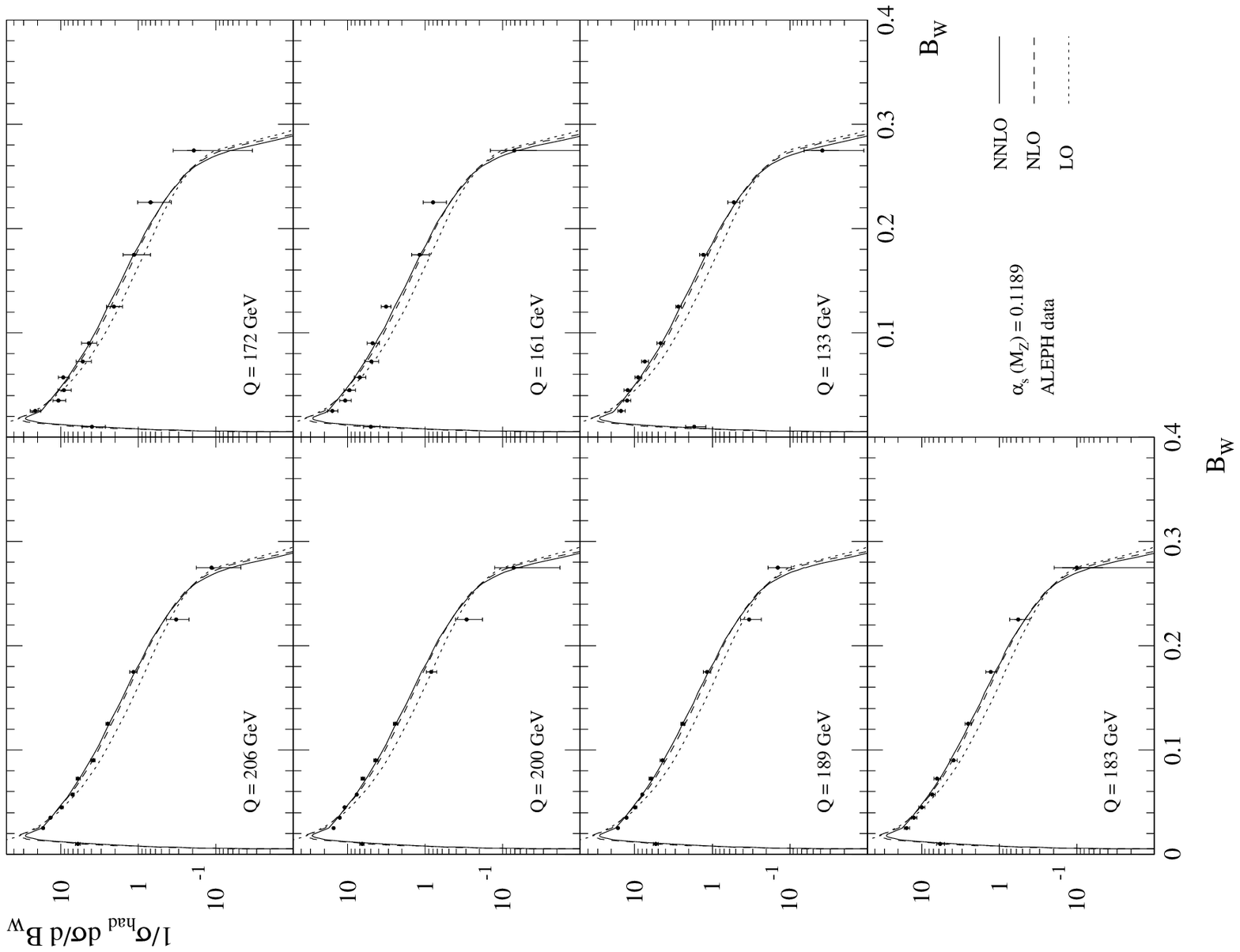,angle=-90,width=15cm}
\end{center}
\caption{Wide jet broadening distribution (with $\mu=Q$ and $\alpha_s(M_Z) = 0.1189$)
at LO (dotted), NLO (dashed) and NNLO
(solid) compared to experimental 
data from ALEPH~\protect{\cite{aleph}} for $Q = 133~{\rm GeV},\ldots,206$~GeV.
\label{fig:aleph-bw}}
\end{figure}

Predictions for the total and wide jet broadenings are
displayed in 
Figures~\ref{fig:bt} and \ref{fig:bw}. 
The solid lines represent the prediction at the  physical scale $Q= M_Z$,
while the shaded bands represent the effect of varying the renormalisation scale upwards and
downwards by a factor of 2. We observe that the relative scale uncertainty 
in the $B_T$ ($B_W$) distribution is reduced by about
40\% (50\%) between NLO and NNLO. 

As anticipated from the discussion in section~\ref{sec:broad}, we observe that 
the perturbative corrections are smaller for $B_W$ than for $B_T$.
In the region where perturbation theory is expected to yield reliable results,
$(B_T,~B_W) > 0.05$, we observe an enhancement of (15-20)\% in $B_T$ and 
of (8-12)\% in $B_W$.  As with $(1-T)$ and the heavy jet mass, the 
two broadenings are identical at leading order, but display a 
largely different 
behaviour in the higher perturbative corrections. 
At smaller values of broadening, large logarithmic
corrections occur which must be resummed~\cite{bwbt}.

To guide the eye, Figures~\ref{fig:bt} and \ref{fig:bw} 
also show hadron-level data from the ALEPH
collaboration~\cite{aleph}.  For both broadenings, we see that the 
NNLO prediction lies closer to the
data, and, in fact, accounts for much of the difference between the NLO 
prediction and the hadron-level data.

The experiments at LEP also gathered data at higher $Q$ values;
Figures~\ref{fig:aleph-bt} and \ref{fig:aleph-bw} compare the parton-level
prediction at $Q = 133~{\rm GeV},\ldots,206$~GeV with hadron-level 
data from the ALEPH
collaboration~\cite{aleph}.
We observe that for all $Q$ values, shape
and normalisation of the parton level NNLO prediction agrees better with the
data than at NLO. 

\subsection{$C$-parameter}

\begin{figure}[t]
\begin{center}
\epsfig{file=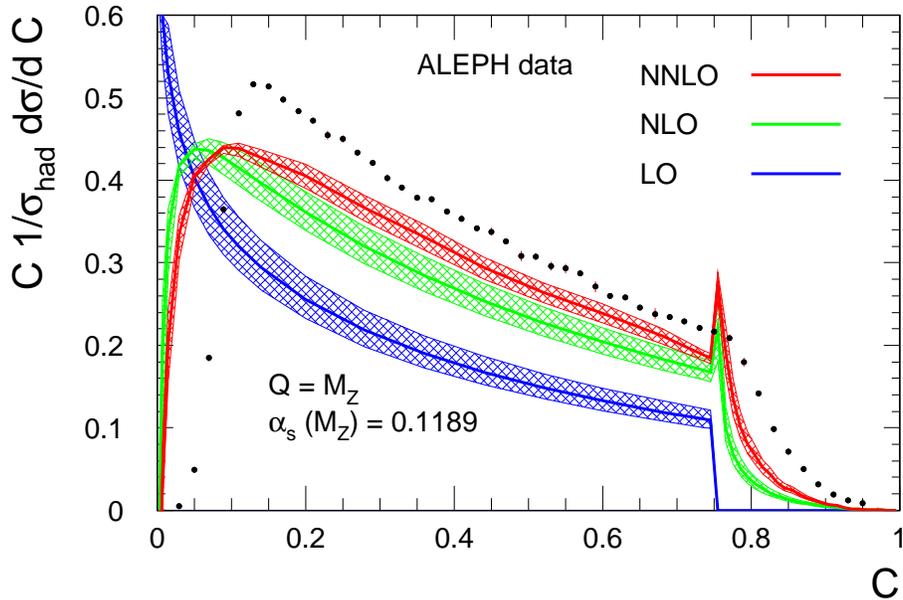,angle=-90,width=12cm}
\end{center}
\caption{C parameter distribution at $Q= M_Z$ at LO (blue), NLO (green) and NNLO
(red). The solid lines represent the prediction for renormalisation scale 
$\mu=Q$ and $\alpha_s(M_Z) = 0.1189$, while the
shaded region shows the variation due to varying
the renormalisation scale between $\mu=Q/2$ and $\mu = 2 Q$.
The data is taken from~\protect{\cite{aleph}}.\label{fig:cpar}}
\end{figure}
\begin{figure}[t]
\begin{center}
\epsfig{file=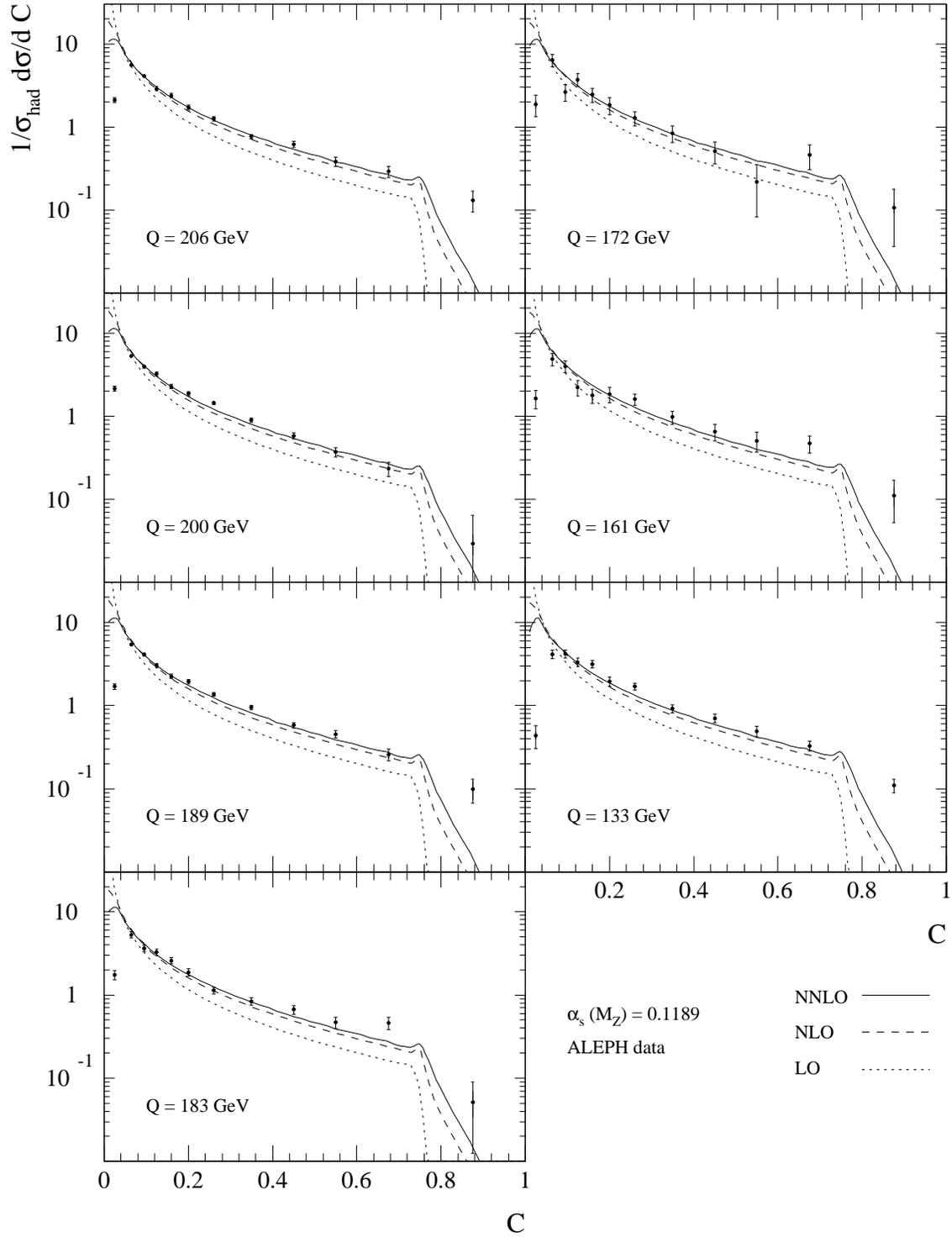,angle=-90,width=15cm}
\end{center}
\caption{C parameter distribution (with $\mu=Q$ and $\alpha_s(M_Z) = 0.1189$)
at LO (dotted), NLO (dashed) and NNLO
(solid) compared to experimental 
data from ALEPH~\protect{\cite{aleph}} for $Q = 133~{\rm GeV},\ldots,206$~GeV.
\label{fig:aleph-cpar}}
\end{figure}

The $C$ parameter is one of the classic event shape observables and we display the perturbative
prediction in  Figure~\ref{fig:cpar}.   The solid lines represent the prediction at the  physical
scale $Q= M_Z$, while the shaded bands represent the effect of varying the renormalisation scale
upwards and downwards by a factor of 2. We observe that the relative scale uncertainty is reduced
by about 40\% between NLO and NNLO.  The NNLO corrections enhance the $C$ parameter
distribution by around (12-20)\% over the  
range $0.1 < C < 0.75$, where $\ln(1/C)$ is not too large.
Figure~\ref{fig:cpar} also shows hadron-level data from 
the ALEPH collaboration~\cite{aleph} and we observe that
the NNLO parton-level 
prediction lies significantly closer to the data, and in fact, 
accounts for about one
third of the difference between the NLO prediction and the data. 

At small $C$,  one expects large logarithmic 
contributions $\ln(1/C)$ appearing in higher
perturbative orders,  thus necessitating an all-orders resummation 
of these logarithmic 
terms~\cite{resum-cpar}.  There are also large logarithms 
around $C \sim 0.75$, due to soft gluon
divergences within the physical region (producing a so-called 
Sudakov shoulder in the distribution) which must also
be resummed~\cite{resum-cpar75} to all orders.

Figure~\ref{fig:aleph-cpar} shows the prediction for a range of $Q$ values together  with the
hadron-level data from the ALEPH collaboration~\cite{aleph}.  For all $Q$ values, the shape and
normalisation of the parton level NNLO prediction agrees slightly  better with the data than at
NLO.

\begin{figure}[th]
\begin{center}
\epsfig{file=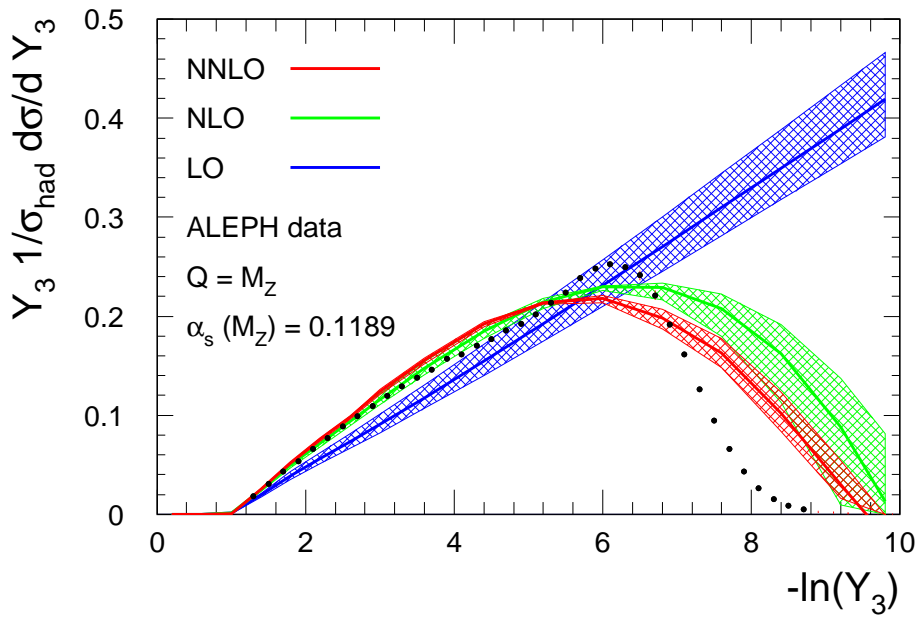,angle=-90,width=12cm}
\end{center}
\caption{
The distribution for the jet transition variable, $Y_3$
at $Q= M_Z$ at LO (blue), NLO (green) and NNLO
(red). The solid lines represent the prediction for renormalisation scale 
$\mu=Q$ and $\alpha_s(M_Z) = 0.1189$, while the
shaded region shows the variation due to varying
the renormalisation scale between $\mu=Q/2$ and $\mu = 2 Q$.
The data is taken from~\protect{\cite{aleph}}.\label{fig:y23}}
\end{figure}
\subsection{$Y_3$}
\begin{figure}[th]
\begin{center}
\epsfig{file=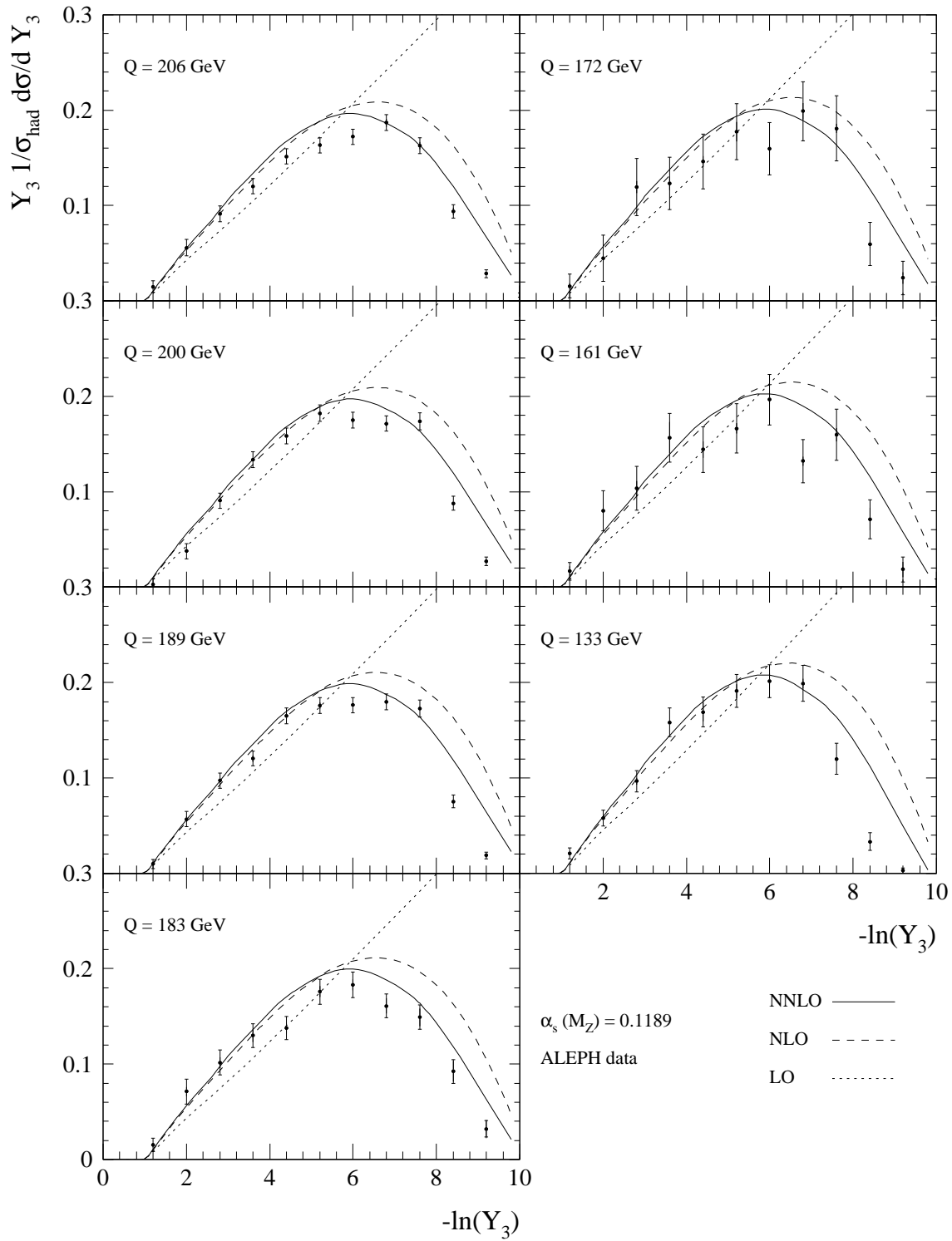,angle=-90,width=15cm}
\end{center}
\caption{The $Y_3$ distribution (with $\mu=Q$ and $\alpha_s(M_Z) = 0.1189$)
at LO (dotted), NLO (dashed) and NNLO
(solid) compared to experimental 
data from ALEPH~\protect{\cite{aleph}} for $Q = 133~{\rm GeV},\ldots,206$~GeV.
\label{fig:aleph-y23}}
\end{figure}
Figure~\ref{fig:y23} displays the perturbative  expression for the $Y_3$
distribution at LO, NLO and NNLO, evaluated  at $Q= M_Z$. The error band indicates
the variation of the  prediction under shifts of the renormalisation scale  in the
range $\mu \in [Q/2;2\,Q]$ around the $e^+e^-$ centre-of-mass  energy $Q$. 
The relative scale uncertainty is reduced by about 50\% between NLO and NNLO. 

The NLO and NNLO corrections change the shape of the distribution 
considerably and introduce a
turnover at $-\ln(Y_3) \sim 5-6$.    We observe that the NNLO corrections 
modify the $Y_3$
distribution by around (3-5)\% over the  range $2 < -\ln(Y_3)< 6$, 
where $-\ln(Y_3)$ is moderate.
At larger $-\ln(Y_3)$,  one does not expect fixed-order perturbation theory
to yield
reliable results and the  large infrared logarithms of the type 
$\alpha_s^n\ln^m(Y_3)$ must be resummed~\cite{resum-y23}. 

The perturbative parton-level prediction is compared with the hadron-level data from the ALEPH
collaboration~\cite{aleph} in Figure~\ref{fig:y23} and  
Figure~\ref{fig:aleph-y23}.  We see that
the quality of the agreement between fixed order perturbation theory and data
 is much more $Q$ dependent than for the other 
observables considered in this paper.      At $Q = M_Z$, the data is
much more sharply peaked than the NNLO prediction. However, at higher $Q$ values shown in
Figure~\ref{fig:aleph-y23}, the agreement between the NNLO prediction and the data around the
peak region is very good - and significantly better than at NLO.

\section{Conclusions and Outlook}
\label{sec:conclusions}

The main goal of this paper is to provide improved 
 theoretical predictions for hadronic event
shape observables in electron-positron annihilation.  To this end, 
we have presented new results
for the next-to-next-to-leading order contributions to a number of important
three-jet-like event shape
observables in $e^+e^-$ collisions. 
These results are obtained using a numerical
program, that is based on the 
matrix elements for $\gamma^* \to 3$ partons at
two-loop,  $\gamma^* \to 4$~partons at one-loop and 
$\gamma^* \to 5$~partons at tree-level. Each
of these contributions becomes singular when 
one or more partons are soft and/or collinear, and
we have developed and implemented an NNLO subtraction formalism to subtract these singularities,
thereby yielding a finite NNLO prediction.  The resulting numerical program, {\tt EERAD3},  yields
the full kinematical information on the partonic final state and can be applied to generic
infrared safe three-jet observables.  

For the six event shapes considered here, and 
in kinematical regions where infrared logarithms are small enough to
render their resummation unnecessary,  
the NLO corrections are generally large - of the order 
30-60\%.  The NNLO effects produce a further 5-20\% correction. 
Comparisons with existing data
from LEP indicate an improved agreement 
with hadronic data and the fixed order NNLO parton-level prediction.
In addition, the remaining theoretical uncertainty estimated by 
varying the renormalisation scale
by a factor of two around the physical scale is also significantly reduced,
typically by 30-50\%.
Importantly, the size of the corrections is different for 
different observables.

Our results for the NNLO corrections open up a whole 
new range of possible 
comparisons with the LEP data. For meaningful comparisons, 
one has to account for hadronisation effects, either by 
introducing hadron-level to parton-level correction factors, or by 
including power-suppressed hadronisation effects in the theoretical 
description. 
A first direct determination of the strong coupling constant from a fit of
next-to-next-to-leading order QCD predictions to event-shape 
variables over a range of $Q$
values will be reported in a separate 
publication~\cite{aspaper} and should yield a much more 
precise value of
$\alpha_s(M_Z)$ than that previously extracted from event shapes. 
  Our predictions can be further improved by a NLL 
resummation of the large
infrared logarithms that are present as $y \to 0$.   The ingredients 
for $\ln R$ matching to NNLO
are available in~\cite{ctw}. Studies 
in this direction are in progress and should yield
a further improvement on the measurement of $\alpha_s(M_Z)$. 
Similarly, our results will also allow a renewed study of
power corrections, now matched to NNLO.

\section*{Acknowledgements}
We would like to thank  G\"unther Dissertori and Hasko Stenzel for 
sharing their expertise on QCD event shapes in many useful discussions.

Part of this work was carried out while the authors were attending 
the programme ``Advancing Collider Physics: From 
Twistors to Monte Carlos'' of the 
Galileo Galilei Institute for
Theoretical Physics (GGI) in Florence. We thank the GGI
 for its hospitality and the 
  Istituto Nazionale di Fisica Nucleare (INFN) for partial support.

This research was supported in part by the Swiss National Science Foundation
(SNF) under contract 200020-117602,
 by the UK Science and Technology Facilities Council and 
 by the European Commission's
 Marie-Curie Research Training Network under contract
MRTN-CT-2006-035505 ``Tools and Precision Calculations for Physics Discoveries
at Colliders''.

\section*{Erratum added}

Our implementation of  NNLO corrections to three-jet-like 
observables~\cite{our3j}, 
which was used for the present work,
 was checked by two subsequent studies: 
the calculation of all logarithmically enhanced contributions to the 
thrust distribution by Becher and Schwartz~\cite{becher}, and an 
independent implementation of our subtraction formulae by 
Weinzierl~\cite{weinzierl}. 

These works uncovered numerical discrepancies in the two-jet limit 
of the observables 
in two of the six colour factor contributions: $N^2$ and $N^0$. 
In~\cite{weinzierl}, it was shown that the origin of these discrepancies is
in an oversubtraction of large-angle soft radiation. We described the
corrected treatment of these terms in the erratum to~\cite{our3j}. 

As a consequence, the 
numercial values of the NNLO coefficients of all event shape distributions 
were modified. The new coefficients are displayed 
by the solid lines in Figures 1--6, our original results are displayed there 
for comparison as dotted lines.  
It can be seen that in the genuine three-jet
region, which is relevant for precision phenomenology, the changes have a
minor numerical impact. The modification to the 
full event shape distributions is too small to be visible, except 
for $Y_3$ in the deep two-jet region. We therefore 
refrain from presenting revised figures 7-18.

\bibliographystyle{JHEP}
 
\end{document}